\documentclass[aps,prb,twocolumn,showpacs,superscriptaddress]{revtex4-2}

\usepackage{graphicx}  % needed for figures
\usepackage{dcolumn}   % needed for some tables
\usepackage{bm}        % for math
\usepackage{enumerate}
\usepackage{amsmath}
\usepackage{esint}
\usepackage{xfrac}
\usepackage{siunitx}
\usepackage{color}
\usepackage{placeins}
\usepackage{float}
\usepackage{caption}
\usepackage{hyperref}
\usepackage{natbib}

\allowdisplaybreaks
\newcommand{\vect}[1]{\mathbf{#1}}
\DeclareMathAlphabet\mathbfcal{OMS}{cmsy}{b}{n}

\begin{document}

\title{$\mathcal{T}$-matrix method for computation of second-harmonic generation upon optical wave scattering from clusters of arbitrary particles: Application to nonlinear optical interaction of bound-states in the continuum}

\date{\today}

\author{Ivan Sekulic}
\email{i.sekulic@ucl.ac.uk}
\affiliation{Department of Electronic and Electrical Engineering, University College London, Torrington Place, London WC1E 7JE, United Kingdom}

\author{Ji Tong Wang}
\email{jitong.wang@ucl.ac.uk}
\affiliation{Department of Electronic and Electrical Engineering, University College London, Torrington Place, London WC1E 7JE, United Kingdom}
\affiliation{Wuzhen Laboratory, EGO Wuzhen Digital Economy Industrial Park, No. 925 Daole Road, Tongxiang City, China}

\author{Jian Wei You}
\email{jvyou@seu.edu.cn}
\affiliation{School of Information Science and Engineering, Southeast University, Nanjing, Jiangsu, China}
	
\author{Nicolae C. Panoiu}
\email{n.panoiu@ucl.ac.uk}
\affiliation{Department of Electronic and Electrical Engineering, University College London, Torrington Place, London WC1E 7JE, United Kingdom}

\begin{abstract}
We derive the $\mathcal{T}$-matrix formalism tailored for numerical analysis of second-harmonic (SH) generation from arbitrarily-shaped particles made of centrosymmetric optical materials. First, the transfer matrix of a single particle is computed \textit{via} the extended boundary condition method, in which the electromagnetic fields both at fundamental frequency and SH are expanded in vector spherical wave functions and the integral formulation is satisfied away from the surface of the scatterer. We allow for the accurate physical description of the SH sources by taking into account both local surface and nonlocal bulk polarization contributions to the nonlinear polarization density source responsible for the generation of the SH signal by a particle. This single-particle formalism is then extended to arbitrary distributions of particles by incorporating into the formalism linear and nonlinear electromagnetic wave scattering from the particles in the cluster. Importantly from a practical point of view, our method can be applied to particles of arbitrary shape made of optical materials characterized by general frequency-dispersion relations, so that it can describe the linear and nonlinear optical response of clusters of metallic, semiconductor, or polaritonic particles, as well as mixtures of such particles. The approach proposed here is faster and more memory-efficient than well-established numerical techniques, especially in the analysis of spheroidal particles, due to the favourable symmetries of spherical wave basis functions used in the wave scattering analysis.

\end{abstract}
\maketitle

\section{Introduction}\label{INTR}
Fast and accurate full-wave computational analysis of resonant electromagnetic response of nanoparticles is of great interest to the scientific community, chiefly due to their widespread use in many areas of science and technology, including plasmonic nanoantennas, remote sensing, optical metamaterials, biology, nonlinear optics, imaging, and colloidal chemistry \cite{Homola,Pendry,Liu,Ziolkowski,Hatab,Butet,Panoiu,Boyd,Ray}, just to name a few. Modern nanofabrication tools, such as electron beam lithography, have made possible to produce particles of virtually any shape, thus the ability to accurately analyze the electromagnetic properties of nanoparticles with different morphologies is a prerequisite in the design process of certain modern optical systems. These nanoparticles are often used as building blocks of metamaterials, so-called \textit{meta-atoms}, and can be shaped in a variety of forms, such as crosses, split-rings, cuboids, spheroids, ellipsoids, \textit{etc}. \cite{Canfield,Valev}. Combined with deep-neural networks, fast numerical solvers have dramatically sped up the computer-aided design phase of the final structures \cite{Malkiel,Molesky,Kudyshev}.

Second-harmonic generation (SHG) is perhaps the most ubiquitous nonlinear optical process encountered in modern photonics. It consists of the generation of waves with frequency twice as large as that of an excitation (pump) optical wave that interacts with a nonlinear optical medium \cite{Boyd,Shen}. The full-wave analytical solutions of this nonlinear electromagnetic problem are restricted to a very few canonical geometries, such as infinite cylinders and spheres \cite{Biris,Pavlyukh,Beer1,Forestiere1,Xu,Sekulic1}. These solutions rely on the expansion of the fundamental-frequency (FF) and second-harmonic (SH) fields into Fourier series of cylindrical or spherical harmonic functions and the application of the tangential field boundary conditions at the surface of the particle. In this way, the Fourier series coefficients can be efficiently computed.

Numerical methods become indispensable when trying to go beyond these canonical geometries. The well-known finite-element method (FEM) \cite{Silvester,Jin} and finite-difference time-domain (FDTD) method \cite{Yee,Taflove} have been successfully applied to analyze the optical response at the SH of systems containing a small number of particles. Their drawbacks pertaining to the SHG analysis are inherited from the linear regime and can be traced to the discretization of the computational domain and application of absorbing boundary conditions in order to artificially truncate it. On the other hand, electromagnetic integral equations implicitly incorporate the radiation boundary condition at infinity making them particularly appealing candidates for tackling complex scattering problems \cite{Mayergoyz,Hohenester1}.

The Poggio–Miller–Chang–Harrington–Wu–Tsai (PMCHWT) surface integral equation \cite{Poggio,Wu} is one of the most popular descriptions of the optical wave scattering from particles in the linear regime, chiefly due to its accuracy and resonance-free behaviour. It has recently been extended to the analysis of SH processes in centrosymmetric media \cite{Makitalo,Forestiere2}. The PMCHWT integral formulation is defined on the surface of the scatterer, mapping the unknown equivalent surface electric and magnetic currents to the tangential field components. In its classical conforming Galerkin method of moments (MoM) numerical discretization \cite{Harrington}, the lowest order Rao-Wilton-Glisson (RWG) functions \cite{Rao} are used as current expansion and field testing set. These are sub-sectional basis functions defined on the tessellation of the particle surface (mesh), linked to the internal edges and straddling two adjacent triangles. This restriction of the RWG expansion functions to the mesh covering the particle surface implies that a large number of such functions are needed to adequately interpolate the unknown surface currents and consequently leads to MoM impedance matrices of particularly large size. Therefore, intense efforts have been invested into developing new approaches to accelerate execution of MoM-based codes by combining highly parallel and performance-oriented implementations with proficient algorithms aimed at reducing the computational complexity and execution time of the code \cite{Song,Bebendorf}. However, even if one employs state-of-the-art acceleration tools, an efficient scattering analysis of clusters containing more than just a few nanoparticles is seriously impeded by the prohibitively large computational resources that are required.

In many instances of practical interest the particles under investigation exhibit axial symmetry, \textit{e.g.} ellipsoids and cylinders, or the response of a cluster of particles with different shapes can be well approximated with one of an assembly of spherical scatterers. In these cases certain symmetries can be exploited by adopting vector spherical wave functions (VSWFs) as the basis functions for the expansion of fields and currents \cite{Varshalovich,Jackson}. This approach is usually coupled to the extended boundary condition method (EBCM) \cite{Waterman1,Waterman2,Barber}, whereby the surface integral equations are imposed at some distance away from the boundary of the particle, in the null-field region. In this process, a $\mathcal{T}$-matrix of the system of particles relating the Fourier coefficients of the incident and scattered fields is introduced \cite{Mishchenko0,Mishchenko1}.

A key strength of the transfer-matrix method (TMM) consists in the fact that the $\mathcal{T}$-matrices of particles in a cluster depend solely on the intrinsic physical properties of the particles. Therefore, they only need to be computed once for a particular frequency, as they do not depend on the angle and polarization of the incident field, nor on the arrangement of particles in the cluster. This feature makes the TMM uniquely suitable for calculations of certain physical quantities, such as radar cross sections. Furthermore, in the calculation of the cluster response, the multiple scattering (electromagnetic coupling) between nanoparticles is conveniently described using the translation-addition theorem for VSWFs \cite{Cruzan,Mackowski,Stout1}, and the \textit{scattering matrix} of the entire nanoparticle aggregate is readily obtained from the single-particle $\mathcal{T}$-matrices and translation-addition matrices. Due to this key property of the TMM, it is alternatively called the multiple-scattering matrix (MSM) method.

The TMM has been extensively developed and used in the linear scattering regime, including for particles of different shapes, sizes, and electric permittivity (see \cite{Mishchenko1} and references therein). It has been extended to the analysis of the SHG in centrosymmetric media, but so far, only the particular cases of spheres and infinite cylinders have been considered \cite{Biris,Xu,Sekulic1}.

In this article, we present a rigorous computational analysis based on the TMM of optical wave scattering from particles made of centrosymmetric materials and of arbitrary shape. Several examples of axially symmetric particles, where the TMM performs the best, are also presented. In particular, we compare our computational predictions with results derived using analytical methods, where they are applicable, or otherwise with results obtained by employing commercially available software based on the FEM. Our study is structured as follows: in the next section, we describe the physics pertaining to the generation of the optical waves at the SH upon the interaction of optical fields applied at the FF with systems of nanoparticles made of centrosymmetric materials. Then, in section \ref{TMM_sin_par}, we introduce the TMM formalism describing the linear and nonlinear optical wave scattering from a single nanoparticle. We extend this concept to the multiparticle case in section \ref{CLUST}, and illustrate how our method performs by considering a few relevant examples in section \ref{NumRes}. A specific application is considered in section \ref{ApplBIC}, where we demonstrate that the SHG in a cluster of silicon nanoparticles can be greatly enhanced when the cluster is designed to support optical resonances at both the FF and SH. In the last section we summarise the main ideas of this work.

\section{System geometry and the origin of SHG in centrosymmetric media}\label{GEOM}
\begin{figure}[!t]
	\centering
	\includegraphics[width=\columnwidth]{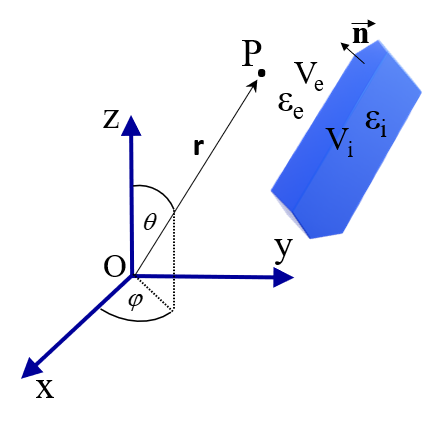}
	\caption{A particle of arbitrary shape made of centrosymmetric optical material described by permittivity $\epsilon_i$ occupying the region $V_i$ embedded in a background medium with permittivity $\epsilon_e$.}\label{fig:Sin_Par}
\end{figure}

The central goal of this study is to develop a framework in which to analyze the SHG upon electromagnetic wave scattering from an arbitrary distribution of $N$ particles made of centrosymmetric, non-magnetic and isotropic optical material. A particle occupies the region $V_i$ and its electrical properties are described by a frequency-dependent complex permittivity function, $\epsilon_{i}$. The particles are embedded in a lossless non-dispersive background medium $V_e$ (usually vacuum) characterized by permittivity $\epsilon_{e}$ (see Fig.~\ref{fig:Sin_Par}). The system is excited by a time-harmonic electromagnetic plane wave with unity amplitude, $\{\vect{E}_{inc}^{\omega},\vect{H}_{inc}^{\omega}\}$, oscillating at FF, $\omega$. Here, we assume an $e^{-i\omega t}$ time-harmonic variation of all fields and sources, but throughout the mathematical derivation of our results this multiplicative factor is omitted. The interior ($V_i$) and the exterior ($V_e$) regions are separated by an orientable surface $\mathcal{S}$ equipped at each surface point with an outwardly oriented unit normal vector, $\hat{\vect{n}}$.

The nonlinear waves generated at the SH oscillate at frequency, $\Omega=2\omega$, and originate from two types of polarization sources. The centrosymmetric crystal lattice of the material forbids the generation of a local dipole polarization source inside the bulk of the constituent medium. However, due to the inversion symmetry breaking at the surface of the particle, a dipole surface polarization density source is induced onto the surface. Although in general this interfacial component of the nonlinear polarization describes accurately enough many of the properties of the experimentally measured SHG signal from metallic particles, for an accurate analysis of nonlinear wave scattering from dielectric particles, a nonlocal polarization source distribution generated inside the bulk of the particle and originating from the induced nonlinear electric quadrupoles and nonlinear magnetic dipoles has to be properly taken into account \cite{Timbrell}.

The nonlinear surface polarization is induced within a few atomic layers just underneath the surface of the particle and is related to the electric field at the FF \textit{via} a third-rank surface susceptibility tensor, $\hat{\bm{\chi}}_s^{(2)}$ \cite{Boyd,Shen}:
\begin{equation}\label{eq:SHSurfPolarization}
\vect{P}_s^{\Omega}(\mathbf{r})=\epsilon_0 \hat{\boldsymbol{\chi}}_s^{(2)}:\vect{E}^{\omega}(\mathbf{r}) \vect{E}^{\omega}(\mathbf{r})\delta(\mathbf{r}-\mathbf{r}_s),
\end{equation}
where $\epsilon_0$ is the vacuum electric permittivity and the Dirac function $\delta(\mathbf{r})$ ensures that the applied electric field is evaluated onto the surface, $\mathbf{r}_{s}\in\mathcal{S}$ (more precisely, the field must be evaluated at a point located on the inner side of the particle surface). The susceptibility tensor $\hat{\boldsymbol{\chi}}_s^{(2)}$ is defined at boundary points $\mathbf{r}_{s}$, and is expressed in a coordinate system defined by a local right-handed orthonormal triplet of vectors $(\hat{\mathbf{n}}, \hat{\mathbf{t}}_{1}, \hat{\mathbf{t}}_{2})$, where $\hat{\mathbf{t}}_{1}$ and $\hat{\mathbf{t}}_{2}$ are unit vectors tangential to the surface at $\mathbf{r}_{s}$. Assuming isotropic surface symmetry, only three tensor components are nonzero and independent: $\chi_{s,\perp\perp\perp}^{(2)}$, $\chi_{s,\perp\parallel\parallel}^{(2)}$, and $\chi_{s,\parallel\perp\parallel}^{(2)} = \chi_{s,\parallel\parallel\perp}^{(2)}$ \cite{Shen}, where the symbols $\perp$ and $\parallel$ refer to the directions perpendicular and parallel to the boundary, respectively. Taking this into account, $\vect{P}_s^{\Omega}(\mathbf{r}_s)$ can be expressed as \cite{Dadap}:
\begin{align}\label{eq:SHSurfPolarization_Comp}
\vect{P}_s^{\Omega}&(\mathbf{r}_s)=\epsilon_0\left\{\hat{\mathbf{n}} \left[\chi_{s,\perp\perp\perp}^{(2)}E_n^{\omega}(\mathbf{r}_s)E_n^{\omega}(\mathbf{r}_s)\right.\right.\\
&+\left.\left.\chi_{s,\perp\parallel\parallel}^{(2)}E_t^{\omega}(\mathbf{r}_s)E_t^{\omega}(\mathbf{r}_s)\right]
+2\chi_{s,\parallel\perp\parallel}^{(2)}E_n^{\omega}(\mathbf{r}_s)\vect{E}_t^{\omega}(\mathbf{r}_s) \right\},\notag
\end{align}
where $\vect{E}_n^{\omega}(\mathbf{r}_s)$ and $\vect{E}_t^{\omega}(\mathbf{r}_s)$ are normal and tangential components of the linear field $\vect{E}^{\omega}(\mathbf{r})$ taken at the inner side of the surface $\mathcal{S}$, at point $\mathbf{r}_s$, \textit{i.e.}
\begin{subequations}\label{eq:EtanEnorm}
	\begin{align}	
	\vect{E}_n^{\omega}(\mathbf{r}_s) &=\left[\hat{\mathbf{n}} \cdot \vect{E}^{\omega}(\mathbf{r})|_{\mathbf{r} = \mathbf{r}_s}\right]\hat{\mathbf{n}}, \label{eq:Enorm}\\
    \vect{E}_t^{\omega}(\mathbf{r}_s) &= - \hat{\mathbf{n}} \times \hat{\mathbf{n}} \times \vect{E}^{\omega}(\mathbf{r})|_{\mathbf{r} = \mathbf{r}_s}.	\label{eq:Etan}
	\end{align}
\end{subequations}

The nonlocal bulk polarization is induced in the interior of the particle by the fundamental field and its spatial derivatives, and is described by the following expression:
\begin{equation}\label{eq:BulkPolar}
\vect{P}_b^{\Omega}(\mathbf{r})=\epsilon_0 \hat{\boldsymbol{\chi}}_b^{(2)} \vdots \vect{E}^{\omega}(\mathbf{r})
\nabla\vect{E}^{\omega}(\mathbf{r}), \quad \vect{r} \in V_i,
\end{equation}
where $\hat{\boldsymbol{\chi}}_b^{(2)}$ is a fourth-rank bulk susceptibility tensor. This polarization can be alternatively expressed as \cite{Bloembergen}:
\begin{equation}
	\vect{P}_b^{\Omega} = \epsilon_0 \left[\beta \vect{E}^{\omega} \nabla \cdot \vect{E}^{\omega}  +
	\gamma \nabla (\vect{E}^{\omega} \cdot \vect{E}^{\omega}) + \delta'
	(\vect{E}^{\omega} \cdot \nabla)\vect{E}^{\omega}\right], \label{eq:BulkPolarSucc}
\end{equation}
with $\beta, \gamma$, and $\delta'$ being material parameters. The first term in this equation vanishes because we assume that the particles are made of homogeneous material so there is no induced charge inside the particles, \textit{i.e.}, $\nabla \cdot \vect{E}^{\omega}=0$. The relative magnitude of the third term as compared to the second one is a question that is still debated. Thus, the contribution of the third term increases when the field at the FF inside the particle becomes strongly inhomogeneous, as it is the case around steep variations of the morphology of the scatterer or nearby sharp corners where so-called \textit{hot spots} can form. Since here we are mostly concerned with axially symmetric and smooth particles, rapid variations of the linear field are not expected and therefore we set $\delta'=0$ as in fact most authors do \cite{Sipe1}. It should be noted, however, that the term proportional to $\delta^{\prime}$ can be accounted for in a rigorous description of generation of SH upon optical wave scattering from particles, but the mathematical description can become quite intricate even in the simple case of a single particle \cite{Forestiere1}. Finally, the total nonlinear polarization density source is equal to the sum of the surface and bulk components:
\begin{equation}\label{eq:TotalPolar}
\vect{P}^{\Omega}(\mathbf{r}) = \vect{P}_b^{\Omega}(\mathbf{r}) + \vect{P}_s^{\Omega}(\mathbf{r}),\quad \vect{r} \in V_i.
\end{equation}
	
The numerical and analytical methods used in the treatment of the SH signal are almost exclusively developed within the so-called undepleted-pump approximation. In this framework, it is assumed that no back-coupling between the nonlinear and linear fields occurs, \textit{i.e.}, the down-conversion process is disregarded. This assumption is valid in most situations of practical interest due to the very low energy conversion efficiency of the SHG process.

\section{The transfer-matrix method for single particle scattering}\label{TMM_sin_par}
In this section, we introduce the mathematical apparatus pertaining to the EBCM, both at the FF and SH. More specifically, we discretize the associated integral operators in a basis consisting of VSWFs and establish a relation between the Fourier coefficients of the series expansions in this basis of the scattered and excitation fields. Importantly, this relation defines the $\mathcal{T}$-matrix of the particle, which is at the heart of the TMM.

\subsection{Linear scattering regime}\label{LinearReg}
The linear part of our TMM, which is used to determine the FF distribution around an arbitrarily shaped particle, is constructed in a well-known manner described in detail in the literature (see, for example, Ref.~\cite{Tsang}). For the sake of completeness and to facilitate the presentation of the nonlinear part of our numerical method, here we will briefly outline the main steps of the derivation, whereas the explicit formulae for the $\mathcal{T}$-matrix elements are given in Appendix~A.

We start by invoking the surface equivalence principle, which divides the original physical problem into two problems equivalent to the initial one. The first problem is concerned with the exterior domain $V_e$ and states that the fields in this region are uniquely determined by the rotated tangential traces of the total fields located at the boundary $\mathcal{S}=\partial V_i$, the equivalent electric and magnetic currents defined as $\vect{J}_{e}^{\omega} = \hat{\mathbf{n}}' \times \vect{H}_{e}^{\omega}$ and $\vect{M}_{e}^{\omega} = -\hat{\mathbf{n}}' \times \vect{E}_{e}^{\omega}$, and the incident fields $\left\{\vect{E}_{inc}^{\omega}(\vect{r}), \vect{H}_{inc}^{\omega}(\vect{r})\right\}$. These (non-physical) equivalent currents radiate outwardly in $V_e$ the scattered fields $\left\{\vect{E}_{s}^{\omega}(\vect{r}), \vect{H}_{s}^{\omega}(\vect{r})\right\}$ and exactly cancel the incident fields inside $V_i$, so as the equalities $\vect{E}_{e}^{\omega}(\vect{r}) = \vect{E}_{s}^{\omega}(\vect{r}) + \vect{E}_{inc}^{\omega}(\vect{r})$ and $\vect{H}_{e}^{\omega}(\vect{r}) = \vect{H}_{s}^{\omega}(\vect{r}) + \vect{H}_{inc}^{\omega}(\vect{r})$ hold throughout the domain $V_e$. These conditions are mathematically expressed by the Stratton-Chu integral field representations \cite{Stratton}:
\begin{subequations}\label{eq:StrattChu}
\begin{align}		
\oiint_\mathcal{S}&\left\{i\omega\mu_0 \hat{\mathbf{G}}_{e}^{\omega}(\vect{r}, \vect{r}') \cdot  \left[\hat{\mathbf{n}}' \times  \vect{H}_{e}^{\omega}(\vect{r}')\right] + \left[\nabla \times  \hat{\mathbf{G}}_{e}^{\omega}(\vect{r}, \vect{r}')\right]\right. \nonumber\\ &\left.\cdot\left[\hat{\mathbf{n}}' \times  \vect{E}_{e}^{\omega}(\vect{r}')\right]\right\}dS +  \vect{E}_{inc}^{\omega}(\vect{r}) = 0, \quad \vect{r} \in V_i,
\label{eq:StrattChu_a}\\
\oiint_\mathcal{S}&\left\{i\omega\mu_0  \hat{\mathbf{G}}_{e}^{\omega}(\vect{r}, \vect{r}') \cdot  \left[\hat{\mathbf{n}}' \times \vect{H}_{e}^{\omega}(\vect{r}')\right] + \left[\nabla \times  \hat{\mathbf{G}}_{e}^{\omega}(\vect{r}, \vect{r}')\right]\right. \nonumber\\ &\left.\cdot\left[\hat{\mathbf{n}}' \times \vect{E}_{e}^{\omega}(\vect{r}')\right]\right\}dS  = \vect{E}_{s}^{\omega}(\vect{r}), \quad \vect{r} \in V_e,\label{eq:StrattChu_b}
\end{align}
\end{subequations}	
where the primed coordinates refer to the integration points on $\mathcal{S}$. According to the first equivalent problem, the electromagnetic medium filling $V_i$ is irrelevant for the field solution in $V_e$, so that we can in principle replace it, to our convenience, with the same background material occupying $V_e$. This way we can use the dyadic Green's function of the infinite space, $\hat{\mathbf{G}}_{e}^{\omega}(\vect{r}, \vect{r}')$, defined as:
\begin{equation}\label{eq:DyadGreen}
\hat{\mathbf{G}}_{e}^{\omega}(\vect{r}, \vect{r}') = \left[\hat{\mathbf{I}} + \frac{\nabla \nabla}{(k_e^{\omega})^2} \right] \frac{e^{ik_e^{\omega}\vert\vect{r}-\vect{r}' \vert}}{4\pi\vert\vect{r}-\vect{r}' \vert},
\end{equation}
where $\hat{\mathbf{I}}$ is the unit dyad and $k_e^{\omega} = \omega\sqrt{\epsilon_e\mu_0}$ is the wave number in the background medium at the FF. The integral field representations \eqref{eq:StrattChu} form the basis of EBCM. Since the equivalent currents radiate in $V_i$ fields that exactly cancel the incident fields, these equations are sometimes called the \textit{null-field} or \textit{extinction} equations. A similar set of electric field integral equations can be derived for the interior region, by exploiting the second equivalent problem. Then, by invoking the duality principle, the magnetic field integral counterparts of Eqs.~\eqref{eq:StrattChu} can be obtained, too \cite{Chew}.

The electric fields incident, scattered, and internal to the particle are expanded in VSWFs \cite{Varshalovich,Jackson}:
\begin{subequations}\label{eq:ExpVSWFE}
\begin{align}
&\vect{E}_{inc}^{\omega}(\vect{r}) = \sum_{\nu\geq1} \left[ q_{\nu}^{\omega}
\vect{M}_{\nu}^{(1)} (k_e^\omega \mathbf{r}) + p_{\nu}^{\omega} \vect{N}_{\nu}^{(1)} (k_e^\omega
\mathbf{r})\right], \quad \vect{r} \in V_e, \label{eq:IncidentVSWFE} \\
&\vect{E}_{s}^{\omega}(\vect{r}) =
\sum_{\nu\geq1} \left[ b_{\nu}^{\omega} \vect{M}_{\nu}^{(3)}
(k_e^\omega \mathbf{r}) + a_{\nu}^{\omega} \vect{N}_{\nu}^{(3)} (k_e^\omega \mathbf{r})\right], \quad \vect{r} \in V_e, \label{eq:ScatteredVSWFE} \\
&\vect{E}_{i}^{\omega}(\vect{r}) = \sum_{\nu\geq1} \left[ c_{\nu}^{\omega}
\vect{M}_{\nu}^{(1)} (k_{i}^\omega \mathbf{r}) + d_{\nu}^{\omega} \vect{N}_{\nu}^{(1)}
(k_{i}^\omega \mathbf{r})\right],\quad \vect{r} \in V_i, \label{eq:InsideVSWFE}
\end{align}
\end{subequations}
where $k_i^{\omega}=\omega\sqrt{\epsilon_i^{\omega}\mu_0}$ is the wave number at the FF of the medium filling the particle. The incident and the internal fields are expanded in a basis of regular spherical functions $\{\vect{M}_{\nu}^{(1)},\vect{N}_{\nu}^{(1)}\}$, which are finite at the origin, the corresponding expansion coefficients being $\{q_{\nu}^{\omega},p_{\nu}^{\omega}\}$ and $\{c_{\nu}^{\omega},d_{\nu}^{\omega}\}$, respectively. The scattered field is expanded using radiating (outgoing) VSWFs, $\{\vect{M}_{\nu}^{(3)},\vect{N}_{\nu}^{(3)}\}$, which satisfy the Silver-M\"{u}ller radiation boundary condition, the Fourier coefficients in this case being $\{b_{\nu}^{\omega},a_{\nu}^{\omega}\}$. Furthermore, the summation over the multi-index $\nu$ combines summations over the orbital and azimuthal indices $l$ and $m$, respectively, so that the summation over this index should be understood as $\sum_{\nu\geq1} \equiv \sum_{l\geq1}\sum_{m=-l}^{l}$. Since the incident field is known, the corresponding Fourier coefficients are known physical quantities. In particular, they are given in Appendix~B for the specific case of an incident linearly polarized plane wave. On the other hand, the expansion coefficients of the internal and scattered fields are unknown quantities and their calculation is the main goal of the TMM.

The expanded incident, scattered, and internal magnetic fields oscillating at frequency $\omega$ can be obtained directly from the electric fields \eqref{eq:ExpVSWFE}, by  dint of Maxwell-Faraday equation for harmonic fields, $\nabla\times\mathbf{E}=i\omega\mathbf{B}$:
\begin{subequations}\label{eq:ExpVSWFH}
\begin{align}
&\vect{H}_{inc}^{\omega}(\vect{r}) = \frac{1}{i Z_e} \sum_{\nu\geq1} \left[ p_{\nu}^{\omega}
\vect{M}_{\nu}^{(1)} (k_e^\omega \mathbf{r}) + q_{\nu}^{\omega} \vect{N}_{\nu}^{(1)} (k_e^\omega
\mathbf{r})\right],\; \vect{r} \in V_e, \label{eq:IncidentVSWFH} \\
&\vect{H}_{s}^{\omega}(\vect{r}) = \frac{1}{i Z_e}
\sum_{\nu\geq1} \left[ a_{\nu}^{\omega} \vect{M}_{\nu}^{(3)}
(k_e^\omega \mathbf{r}) + b_{\nu}^{\omega} \vect{N}_{\nu}^{(3)} (k_e^\omega \mathbf{r})\right], \; \vect{r} \in V_e, \label{eq:ScatteredVSWFH} \\
&\vect{H}_{i}^{\omega}(\vect{r}) = \frac{1}{i Z_i^{\omega}} \sum_{\nu\geq1} \left[ d_{\nu}^{\omega}
\vect{M}_{\nu}^{(1)} (k_{i}^\omega \mathbf{r}) + c_{\nu}^{\omega} \vect{N}_{\nu}^{(1)}
(k_{i}^\omega \mathbf{r})\right].\; \vect{r} \in V_i, \label{eq:InsideVSWFH}
\end{align}
\end{subequations}
In these equations, $Z_e=\sqrt{\sfrac{\mu_0}{\epsilon_e}}$ and $Z_i^{\omega}=\sqrt{\sfrac{\mu_0}{\epsilon_i^{\omega}}}$ are the impedances of the background and the medium inside the particle, respectively. In the derivation of Eqs.~\eqref{eq:ExpVSWFH}, we have relied on the curl identities that relate the VSWFs, given in Appendix~B.

The dyadic Green function $\hat{\mathbf{G}}_{e}^{\omega}$ is the fundamental solution of the vector Helmholtz equation in 3D and can be expanded in a sum of dyadic products of VSWFs \cite{Morse}:
\begin{subequations}\label{eq:DyadGreenExp}
\begin{align}	
\hat{\mathbf{G}}_{e}^{\omega}(\vect{r}, \vect{r}')=& ik_e^{\omega} \sum_{\nu\geq1}\left[	\vect{M}_{\bar{\nu}}^{(3)}(k_{e}^\omega \mathbf{r}')\vect{M}_{\nu}^{(1)}(k_{e}^\omega \mathbf{r})\right.\nonumber\\
&\left.+\vect{N}_{\bar{\nu}}^{(3)}(k_{e}^\omega \mathbf{r}')\vect{N}_{\nu}^{(1)}(k_{e}^\omega \mathbf{r})\right], \quad |\vect{r}|<|\vect{r}'|, \label{eq:DyadGreenExp_a}\\
\hat{\mathbf{G}}_{e}^{\omega}(\vect{r},\vect{r}')=& ik_e^{\omega} \sum_{\nu\geq1}\left[	\vect{M}_{\bar{\nu}}^{(1)}(k_{e}^\omega \mathbf{r}')\vect{M}_{\nu}^{(3)}(k_{e}^\omega \mathbf{r})\right.\nonumber\\
&\left.+\vect{N}_{\bar{\nu}}^{(1)}(k_{e}^\omega \mathbf{r}')\vect{N}_{\nu}^{(3)}(k_{e}^\omega \mathbf{r})\right], \quad |\vect{r}|>|\vect{r}'|, \label{eq:DyadGreenExp_b}
\end{align}
\end{subequations}
where the multi-index $\bar{\nu}$ combines the indices $(l,-m)$.

In the absence of nonlinear optical interactions, the tangent components of the electric and magnetic fields are continuous functions across the surface of the particle, a property that is expressed as:
\begin{subequations}\label{eq:TancondL}
\begin{align}
&\hat{\vect{n}} \times\left[\vect{E}_{e}^{\omega}(\vect{r})-\vect{E}_{i}^{\omega}(\vect{r})\right]=0 ,\quad \vect{r} \in \mathcal{S}\label{eq:TancondL_a}, \\
&\hat{\vect{n}}\times\left[\vect{H}_{e}^{\omega}(\vect{r})-\vect{H}_{i}^{\omega}(\vect{r})\right]=0, \quad \vect{r} \in \mathcal{S}.\label{eq:TancondL_b}
\end{align}
\end{subequations}

Now we turn our attention to Eq.~\eqref{eq:StrattChu_a}, and plug in this integral equation the VSWF-expansions of the internal electric field \eqref{eq:InsideVSWFE}, internal magnetic field \eqref{eq:InsideVSWFH}, the incident electric field \eqref{eq:IncidentVSWFE}, as well as the dyadic Green function \eqref{eq:DyadGreenExp_a} defined at points just inside the particle. Equating the coefficients multiplying regular spherical harmonics $\vect{M}_{\nu}^{(1)}$ and $\vect{N}_{\nu}^{(1)}$ with the same multi-index $\nu$, we obtain a matrix equation relating the Fourier coefficients of the internal and incident fields \cite{Mishchenko0}:
\begin{equation}
	\mathbf{Q}^{\omega(1,3)}\mathbf{h}^{\omega}  = -\mathbf{g}^{\omega}.
	\label{eq:Q13matrixFF}
\end{equation}
In this equation, the (unknown) internal field coefficients $\{c_{\nu}^{\omega},d_{\nu}^{\omega}\}$ are assembled in the vector $\mathbf{h}^{\omega}$, whereas the (known) incident field coefficients $\{q_{\nu}^{\omega},p_{\nu}^{\omega}\}$ are stacked up in the vector $\mathbf{g}^{\omega}$. The superscript $(1,3)$ is used as a reminder that the matrix elements of $\mathbf{Q}^{\omega(1,3)}$ are expressed as combinations of regular and outgoing VSWFs. The specific mathematical expressions of these matrix elements are provided in the Appendix~A.

We now consider the second of Eqs.~\eqref{eq:StrattChu}, from which one can establish a relation between the expansion coefficients of the internal and scattered fields. Analogously to the process just described, we first use the field boundary conditions \eqref{eq:TancondL} to substitute in this equation the external fields with the internal ones, and then insert the expansions of the internal and scattered fields as well as the expansion of the dyadic Green function \eqref{eq:DyadGreenExp_b}. Finally, in the resulting expression we equate the coefficients of the like-terms $\vect{M}_{\nu}^{(3)}$ and $\vect{N}_{\nu}^{(3)}$ that correspond to the same multi-index $\nu$. As a result of these mathematical manipulations we are left with the matrix equation connecting the Fourier expansion coefficients of the internal and scattering fields:
\begin{equation}
	\mathbf{Q}^{\omega(1,1)}\mathbf{h}^{\omega}  = \mathbf{f}^{\omega}.
	\label{eq:Q11matrixFF}
\end{equation}
In this relation, the (unknown) scattering Fourier coefficients at the FF, $\{b_{\nu}^{\omega},a_{\nu}^{\omega}\}$, are collected in the vector $\mathbf{f}^{\omega}$, and the superscript $(1,1)$ in Eq.~\eqref{eq:Q11matrixFF} indicates that the matrix elements of $\mathbf{Q}^{\omega(1,1)}$ contain only regular VSWFs.

Combining Eqs.~\eqref{eq:Q13matrixFF} and \eqref{eq:Q11matrixFF}, we arrive to the $\mathcal{T}$-matrix that maps the Fourier coefficients of the incident field to those of the scattering field,
\begin{equation}
	\mathbf{T}^{\omega}\mathbf{g}^{\omega}  = \mathbf{f}^{\omega},
	\label{eq:TmatrixFF}
\end{equation}
with the $\mathcal{T}$-matrix of the particle given by the product of \textbf{Q}-matrices \cite{Mishchenko0}:
\begin{equation}
	\mathbf{T}^\omega = -\mathbf{Q}^{\omega(1,1)}\left[\mathbf{Q}^{\omega(1,3)}\right]^{-1}.
	\label{eq:TmatrixfromQ}
\end{equation}

Although in principle this formalism can be applied to scatterers of practically any shape, because of the spherical nature of the underlying expansion functions, the TMM is especially suitable for the analysis of optical wave scattering from particles with axial symmetry, such as ellipsoids and cylinders. In these cases, the numerical method converges relatively fast. More precisely, both the near- and far-field physical quantities can be accurately calculated using a fairly small number of VSWFs in the field expansions. Moreover, the VSWFs can be viewed as \textit{entire-domain} basis functions, \textit{i.e.} they are defined irrespective of the surface discretization mesh, so that the constraints pertaining to the mesh generation process can be more relaxed when compared to the MoM. Specifically, the RWG basis functions are defined over two adjacent triangles, so that the underlying surface discretization mesh has to be conformal with all pairs of adjacent facets sharing a certain edge. On the other hand, the surface discretization mesh needed for the evaluation of the elements of the $\mathcal{T}$-matrix can be nonconformal as well, where adjacent triangles may not share single edges. This is a truly significant alleviation on the mesh generation constraints. However, if the shape of the particle is highly irregular, with sharp corners, tips, and deep corrugations, strongly inhomogeneous near-fields are induced. In such cases, MoM could be preferable to TMM since the flexibility provided by the RWG basis functions allows one to accurately interpolate the rapidly varying surface current densities \cite{Sekulic2}.

\subsection{Wave scattering at the second-harmonic}\label{NonLinearReg}
Unlike the linear electromagnetic field, whose sources are not contained in the physical domain of interest, the nonlinear fields are generated by sources (nonlinear polarization) distributed inside the particle or at its surface. Hence, the electromagnetic fields oscillating at the SH frequency, $\Omega$, must satisfy the following \textit{inhomogeneous} Maxwell equations valid inside the particle:
\begin{subequations}\label{eq:MaxwinsSH}
\begin{align}
&\nabla \times \vect{H}_{i}^{\Omega}(\vect{r})+i\Omega
\epsilon_{i}^{\Omega}\vect{E}_{i}^{\Omega}(\vect{r})=-i\Omega\vect{P}_{b}^{\Omega}(\vect{r}),\quad
\vect{r}\in V_{i} \label{eq:MaxwinsSH_a}, \\
&\nabla \times \vect{E}_{i}^{\Omega}(\vect{r})-i\Omega\mu_{0}\vect{H}_{i}^{\Omega}(\vect{r})= 0,
\quad \vect{r}\in V_{i}, \label{eq:MaxwinsSH_b}
\end{align}
\end{subequations}
and the homogeneous system of equations defined in the surrounding medium:
\begin{subequations}\label{eq:MaxwoutSH}
\begin{align}
&\nabla \times \vect{H}_{e}^{\Omega}(\vect{r}) + i\Omega \epsilon_{e}\vect{E}_{e}^{\Omega}(\vect{r})=0, \quad \vect{r}\in V_{e}, \label{eq:MaxwoutSH_a} \\
&\nabla \times \vect{E}_{e}^{\Omega}(\vect{r}) - i\Omega  \mu_{0}\vect{H}_{e}^{\Omega}(\vect{r})=
0, \quad \vect{r}\in V_{e}. \label{eq:MaxwoutSH_b}
\end{align}
\end{subequations}
The external field solutions $\left\{\vect{E}_{e}^{\Omega}, \vect{H}_{e}^{\Omega}\right\}$ at the SH must satisfy in the far-field region the Silver-M\"{u}ller radiation condition, whereas on the boundary of the particle the system of Maxwell equations is supplemented with a set of \textit{nonlinear} tangential field boundary conditions:
\begin{subequations}\label{eq:TancondSH}
\begin{align}
&\hat{\vect{n}}\times\left[\vect{E}_{e}^{\Omega}(\vect{r})-\vect{E}_i^{\Omega}(\vect{r})\right]= -\bm{\mathcal{M}}^{\Omega}, \quad \vect{r}\in \mathcal{S}, \label{eq:TancondSH_a} \\
&\hat{\vect{n}}\times\left[\vect{H}_{e}^{\Omega}(\vect{r})-\vect{H}_i^{\Omega}(\vect{r})\right]= \bm{\mathcal{J}}^{\Omega},
\quad \vect{r}\in \mathcal{S}. \label{eq:TancondSH_b}
\end{align}
\end{subequations}

It can be seen from these equations that, unlike the linear case described by Eqs.~\eqref{eq:TancondL}, the tangent fields at the SH are discontinuous. These field discontinuities are induced by the surface nonlinear polarization, and can be described by introducing nonlinear surface electric and magnetic currents, $\bm{\mathcal{J}}^{\Omega}$ and $\bm{\mathcal{M}}^{\Omega}$, respectively. Specifically, these nonlinear surface currents are related to the surface nonlinear polarization density as \cite{Heinz}:
\begin{subequations}\label{eq:NonlCurr}
\begin{align}
&\bm{\mathcal{J}}^{\Omega}(\vect{r}_s) =  i\Omega \hat{\vect{n}} \times \left[\hat{\vect{n}} \times \bm{\vect{P}}_{s}^{\Omega}(\vect{r}_s)\right], \label{eq:NonlCurr_a}\\
&\bm{\mathcal{M}}^{\Omega}(\vect{r}_s) =\frac{1}{\epsilon_0} \hat{\vect{n}} \times \nabla_{S}
\left[\hat{\vect{n}} \cdot \bm{\vect{P}}_{s}^{\Omega}(\vect{r}_s)\right], \label{eq:NonlCurr_b}
\end{align}
\end{subequations}
where the operator $\nabla_S$ is defined on a plane tangent to the surface at the field evaluation point.

We follow the well-known technique of solving inhomogeneous systems of partial differential equations by seeking the solution as the sum of the general solution of the associated homogeneous system satisfying the homogeneous boundary conditions \eqref{eq:TancondL} and a particular solution of the inhomogeneous system obeying the nonlinear boundary conditions \eqref{eq:TancondSH}. Applying this procedure to the Maxwell equations at the the SH \eqref{eq:MaxwinsSH}, we look for the electromagnetic fields $\left\{\bar{\vect{E}}_{i}^{\Omega}, \bar{\vect{H}}_{i}^{\Omega}\right\}$ satisfying the homogeneous system of equations obtained from Eqs.~\eqref{eq:MaxwinsSH} by setting $\vect{P}_{b}^{\Omega}=0$, and a particular solution $\left\{\vect{E}_{i,p}^{\Omega}, \vect{H}_{i,p}^{\Omega}\right\}$ that satisfies the inhomogeneous system \eqref{eq:MaxwinsSH}.

As can be easily verified by direct substitution into Eqs.~\eqref{eq:MaxwinsSH}, one particular solution to Eqs.~\eqref{eq:MaxwinsSH} is:
\begin{subequations}\label{eq:PartsolSH}
\begin{align}
&\vect{H}_{i,p}^{\Omega}(\mathbf{r}) = 0, \quad \vect{r}\in V_{i},\label{eq:PartSH_H} \\
&\vect{E}_{i,p}^{\Omega}(\mathbf{r}) = -\frac{1}{\epsilon_{i}^{\Omega}}\vect{P}_{b}^{\Omega}(\vect{r}), \quad \vect{r}\in V_{i}\label{eq:PartSH_E}, \end{align}
\end{subequations}
so that the full solution to the system \eqref{eq:MaxwinsSH} is:
\begin{subequations}\label{eq:ComplsolSHint}
\begin{align}
&\vect{H}_{i}^{\Omega}(\mathbf{r}) = \bar{\vect{H}}_{i}^{\Omega}(\mathbf{r}) + \vect{H}_{i,p}^{\Omega}(\mathbf{r}), \quad \vect{r}\in V_{i}, \\
&\vect{E}_{i}^{\Omega}(\mathbf{r}) = \bar{\vect{E}}_{i}^{\Omega}(\mathbf{r}) + \vect{E}_{i,p}^{\Omega}(\mathbf{r}), \quad \vect{r}\in V_{i}.
\end{align}
\end{subequations}

Solutions $\left\{\bar{\vect{E}}_{i}^{\Omega}, \bar{\vect{H}}_{i}^{\Omega}\right\}$ and $\left\{\vect{E}_{e}^{\Omega}, \vect{H}_{e}^{\Omega}\right\}$ of the corresponding homogeneous systems of equations valid inside and outside the particle, respectively, must obey the following boundary conditions on the particle surface, $\mathcal{S}$:
\begin{subequations}\label{eq:TancondSHhom}
\begin{align}
&\hat{\vect{n}} \times[\vect{E}_{e}^{\Omega}(\vect{r})-\bar{\vect{E}}_i^{\Omega}(\vect{r})]=-\bm{\mathcal{M}}^{\Omega}(\vect{r}) + \hat{\vect{n}} \times \vect{E}_{i,p}^{\Omega}(\vect{r}),\quad \vect{r}\in \mathcal{S}, \label{eq:TancondSHhom_a} \\
&\hat{\vect{n}}
\times[\vect{H}_{e}^{\Omega}(\vect{r})-\bar{\vect{H}}_i^{\Omega}(\vect{r})]=\bm{\mathcal{J}}^{\Omega}(\vect{r}),
\quad \vect{r}\in \mathcal{S}. \label{eq:TancondSHhom_b}
\end{align}
\end{subequations}

Analogously to the integral field representations \eqref{eq:StrattChu} derived for the fields at the FF, following the equivalence principle, we introduce the Stratton-Chu integral equations for the SH fields:
\begin{subequations}\label{eq:StrattChuSH}
\begin{align}
\oiint_\mathcal{S}&\left\{ i\Omega\mu_0  \hat{\mathbf{G}}_{e}^{\Omega}(\vect{r}, \vect{r}') \cdot  \left[\hat{\mathbf{n}}' \times \vect{H}_{e}^{\Omega}(\vect{r}')\right] + \left[\nabla \times  \hat{\mathbf{G}}_{e}^{\Omega}(\vect{r}, \vect{r}')\right]\right.\nonumber\\& \left.\cdot \left[\hat{\mathbf{n}}' \times \vect{E}_{e}^{\Omega}(\vect{r}')\right]\right\}dS= 0, \quad \vect{r} \in V_i,\label{eq:StrattChuSH_a}\\
\oiint_\mathcal{S}&\left\{ i\Omega\mu_0  \hat{\mathbf{G}}_{e}^{\Omega}(\vect{r}, \vect{r}') \cdot \left[ \hat{\mathbf{n}}' \times \vect{H}_{e}^{\Omega}(\vect{r}')\right] + \left[\nabla \times  \hat{\mathbf{G}}_{e}^{\Omega}(\vect{r}, \vect{r}')\right]\right.\nonumber\\& \left.\cdot \left[\hat{\mathbf{n}}' \times \vect{E}_{e}^{\Omega}(\vect{r}')\right]\right\}dS  = \vect{E}_{e}^{\Omega}(\vect{r}), \quad \vect{r} \in V_e,\label{eq:StrattChuSH_b}	
\end{align}
\end{subequations}
where $\hat{\mathbf{G}}_{e}^{\Omega}(\vect{r}, \vect{r}')$ is the unbounded dyadic Green function at the SH. It is given by Eq.~\eqref{eq:DyadGreen}, in which we replace $k_e^{\omega}\rightarrow k_e^{\Omega}$. In a similar way to that described by Eqs.~\eqref{eq:DyadGreenExp}, this dyadic Green function can be expanded in VSWFs:
\begin{subequations}\label{eq:DyadGreenExpSH}
\begin{align}	
\hat{\mathbf{G}}_{e}^{\Omega}(\vect{r}, \vect{r}') =& ik_e^{\Omega} \sum_{\nu\geq1}	\left[\vect{M}_{\bar{\nu}}^{(3)}(k_{e}^\Omega \mathbf{r}')\vect{M}_{\nu}^{(1)}(k_{e}^\Omega \mathbf{r})\right.\nonumber\\
&\left.+\vect{N}_{\bar{\nu}}^{(3)}(k_{e}^\Omega \mathbf{r}')\vect{N}_{\nu}^{(1)}(k_{e}^\Omega \mathbf{r})\right], \quad |\vect{r}|<|\vect{r}'|, \label{eq:DyadGreenExpSH_a}\\
\hat{\mathbf{G}}_{e}^{\Omega}(\vect{r}, \vect{r}') =& ik_e^{\Omega} \sum_{\nu\geq1}	\left[\vect{M}_{\bar{\nu}}^{(1)}(k_{e}^\Omega \mathbf{r}')\vect{M}_{\nu}^{(3)}(k_{e}^\Omega \mathbf{r})\right.\nonumber\\
&\left.+\vect{N}_{\bar{\nu}}^{(1)}(k_{e}^\Omega \mathbf{r}')\vect{N}_{\nu}^{(3)}(k_{e}^\Omega \mathbf{r})\right], \quad |\vect{r}|>|\vect{r}'|. \label{eq:DyadGreenExpSH_b}
\end{align}
\end{subequations}

The relations expressed in Eqs.~\eqref{eq:StrattChuSH} are conceptually similar to the corresponding integral equations at the FF, with the operators mapping the rotated tangential traces of the total external SH fields $\{\hat{\mathbf{n}}' \times \vect{E}_{e}^{\Omega}, \hat{\mathbf{n}}' \times \vect{H}_{e}^{\Omega}\}$ to the SH electric field, $\vect{E}_{e}^{\Omega}(\vect{r})$, outside the particle [as per Eq.~\eqref{eq:StrattChuSH_b}], and to the zero field inside it -- see Eq.~\eqref{eq:StrattChuSH_a}. It is important to stress that there are no SH excitation field sources in the exterior region, thus the total fields outside the particle are equal to the \textit{scattered} fields. As we will see later on, the SH excitation sources are introduced in the integral formalism \textit{via} boundary conditions and the particular solution given by Eqs.~\eqref{eq:PartsolSH}.

We now proceed with the expansion of SH electric fields, which are solutions of the homogeneous Maxwell equations at the SH, into Fourier series of VSWFs:
\begin{subequations}\label{eq:ExpVSWFE_SH}
\begin{align}
&\vect{E}_{e}^{\Omega}(\vect{r}) =
\sum_{\nu\geq1} \Big[ b_{\nu}^{\Omega} \vect{M}_{\nu}^{(3)}
(k_e^\Omega \mathbf{r}) + a_{\nu}^{\Omega} \vect{N}_{\nu}^{(3)} (k_e^\Omega \mathbf{r})\Big], \quad \vect{r} \in V_e, \label{eq:ScatteredVSWFE_SH} \\
&\bar{\vect{E}}_{i}^{\Omega}(\vect{r}) =  \sum_{\nu\geq1} \Big[ c_{\nu}^{\Omega}
\vect{M}_{\nu}^{(1)} (k_{i}^\Omega \mathbf{r}) + d_{\nu}^{\Omega} \vect{N}_{\nu}^{(1)}
(k_{i}^\Omega \mathbf{r})\Big],\quad \vect{r} \in V_i, \label{eq:InsideVSWFE_SH}
\end{align}
\end{subequations}
with $k_{e}^\Omega$ and $k_{i}^\Omega$ representing wave numbers associated to the exterior and the interior regions calculated at the SH frequency, respectively. The corresponding Fourier series expansions of the SH magnetic fields are derived from Eqs.~\eqref{eq:ExpVSWFE_SH} combined with Eqs.~\eqref{eq:MaxwinsSH_b} and \eqref{eq:MaxwoutSH_b}, and can be written as:
\begin{subequations}\label{eq:ExpVSWFH_SH}
\begin{align}
&\vect{H}_{e}^{\Omega}(\vect{r}) = \frac{1}{i Z_e}
\sum\limits_{\nu\geq1} \Big[ a_{\nu}^{\Omega} \vect{M}_{\nu}^{(3)}
(k_e^\Omega \mathbf{r}) + b_{\nu}^{\Omega} \vect{N}_{\nu}^{(3)} (k_e^\Omega \mathbf{r})\Big], ~ \vect{r} \in V_e, \label{eq:ScatteredVSWFH_SH} \\
&\bar{\vect{H}}_{i}^{\Omega}(\vect{r}) = \frac{1}{i Z_i^{\Omega}} \sum\limits_{\nu\geq1} \Big[ d_{\nu}^{\Omega}
\vect{M}_{\nu}^{(1)} (k_{i}^\Omega \mathbf{r}) + c_{\nu}^{\Omega} \vect{N}_{\nu}^{(1)}
(k_{i}^\Omega \mathbf{r})\Big],~ \vect{r} \in V_i. \label{eq:InsideVSWFH_SH}
\end{align}
\end{subequations}

Regular spherical harmonics are used in the expansion of the internal SH fields $\left\{\bar{\vect{E}}_{i}^{\Omega}, \bar{\vect{H}}_{i}^{\Omega}\right\}$, as they must be finite at the origin, whereas the external SH fields $\left\{\vect{E}_{e}^{\Omega}, \vect{H}_{e}^{\Omega}\right\}$ are expanded in the basis of outgoing vector spherical harmonics. The Fourier expansion coefficients $\{c_{\nu}^{\Omega},d_{\nu}^{\Omega}\}$ for the internal fields and $\{b_{\nu}^{\Omega},a_{\nu}^{\Omega}\}$ for the external ones are the unknown physical quantities that we need to calculate. These physical quantities will be determined by solving a certain linear system of equations that we will construct in what follows.

In the construction of the $\mathcal{T}$-matrix at the SH, which relates the Fourier expansion coefficients of the excitation and external (scattering) fields, we employ the SH integral relation \eqref{eq:StrattChuSH_a}, which is valid in the interior of the particle, in which we replace the external field traces $\left\{\hat{\mathbf{n}}' \times \vect{E}_{e}^{\Omega}, \hat{\mathbf{n}}' \times \vect{H}_{e}^{\Omega}\right\}$ with the homogeneous internal field traces $\left\{\hat{\mathbf{n}}' \times \bar{\vect{E}}_{i}^{\Omega}, \hat{\mathbf{n}}' \times \bar{\vect{H}}_{i}^{\Omega}\right\}$. This latter step is performed by taking into account the boundary conditions \eqref{eq:TancondSHhom}. After these mathematical manipulations, we arrive at the following integral equation:
\begin{widetext}
\begin{align}\label{eq:TMMSHintegral_Inn}		
\oiint_\mathcal{S}\left\{ i\Omega\mu_0  \hat{\mathbf{G}}_{e}^{\Omega}(\vect{r}, \vect{r}') \cdot  \left[\hat{\mathbf{n}}' \times \bar{\vect{H}}_{i}^{\Omega}(\vect{r}')\right]\right. &+\left.\left[\nabla \times  \hat{\mathbf{G}}_{e}^{\Omega}(\vect{r}, \vect{r}')\right] \cdot \left[\hat{\mathbf{n}}' \times \bar{\vect{E}}_{i}^{\Omega}(\vect{r}')\right]\right\}dS= - i\Omega\mu_0 \oiint_\mathcal{S}   \hat{\mathbf{G}}_{e}^{\Omega}(\vect{r}, \vect{r}') \cdot \bm{\mathcal{J}}^{\Omega}(\vect{r}')dS\nonumber\\& + \oiint_\mathcal{S} \left[\nabla \times  \hat{\mathbf{G}}_{e}^{\Omega}(\vect{r}, \vect{r}')\right] \cdot \left[\bm{\mathcal{M}}^{\Omega}(\vect{r}') - \hat{\mathbf{n}}' \times \vect{E}_{i,p}^{\Omega}(\mathbf{r}')\right]dS, \quad \vect{r} \in V_i,
\end{align}
where the r.h.s. of this equation represents the contribution of the SH polarization sources. Substituting in the equation above the VSWF-expansions of the internal homogeneous fields at the SH, described by Eqs.~\eqref{eq:InsideVSWFE_SH} and \eqref{eq:InsideVSWFH_SH}, as well as the expansion in VSWFs of the dyadic Green function at the SH and  valid for the points inside the particle, given by Eq.~\eqref{eq:DyadGreenExpSH_a}, and equating the coefficients multiplying regular spherical harmonics with the same multi-index, we are left with a linear system of equations represented in the matrix form as:
\begin{equation}
	\mathbf{Q}^{\Omega(1,3)}\mathbf{h}^{\Omega}  = \mathbf{g}^{\Omega(3)}.
	\label{eq:Q13matrixSH}
\end{equation}

The matrix elements $\mathbf{Q}^{\Omega(1,3)}$ are of a form similar to that of the FF matrix elements $\mathbf{Q}^{\omega(1,3)}$, but computed at the frequency $\Omega$. Their specific expressions are given in Appendix~A. Moreover, the unknown Fourier coefficients of the internal homogeneous fields, $\{c_{\nu}^{\Omega},d_{\nu}^{\Omega}\}$, are stacked up in the vector $\mathbf{h}^{\Omega}$. The nonlinear excitation source components are collected in the vector $\mathbf{g}^{\Omega(3)}$ and are defined by surface integrals of products of tangential components of the nonlinear current sources and outgoing VSWFs. The specific formulae of the components of $\mathbf{g}^{\Omega(3)}$ are provided in Appendix~A, too. Note that $\mathbf{g}^{\Omega(3)}$ can be computed once the fields at the FF are determined, so that as far as the calculation of the nonlinear fields is concerned the vector $\mathbf{g}^{\Omega(3)}$ should be viewed as a known physical quantity.

In order to evaluate the Fourier coefficients $\{b_{\nu}^{\Omega},a_{\nu}^{\Omega}\}$ of the external fields at the SH, we use Eq.~\eqref{eq:StrattChuSH_b} from which again the external fields are eliminated with the help of the boundary conditions \eqref{eq:TancondSHhom}. The resulting equation can be cast in the form:
\begin{align}\label{eq:TMMSHintegral_Out}		
\oiint_\mathcal{S}\left\{ i\Omega\mu_0  \hat{\mathbf{G}}_{e}^{\Omega}(\vect{r}, \vect{r}') \cdot  \left[\hat{\mathbf{n}}' \times \bar{\vect{H}}_{i}^{\Omega}(\vect{r}')\right]\right. &+\left.\left[\nabla \times  \hat{\mathbf{G}}_{e}^{\Omega}(\vect{r}, \vect{r}')\right] \cdot \left[\hat{\mathbf{n}}' \times \bar{\vect{E}}_{i}^{\Omega}(\vect{r}')\right]\right\}dS= \vect{E}_{e}^{\Omega}(\vect{r}) - i\Omega\mu_0  \oiint_\mathcal{S}  \hat{\mathbf{G}}_{e}^{\Omega}(\vect{r}, \vect{r}') \cdot \bm{\mathcal{J}}^{\Omega}(\vect{r}')dS\nonumber\\ &+ \oiint_\mathcal{S} \left[\nabla \times  \hat{\mathbf{G}}_{e}^{\Omega}(\vect{r}, \vect{r}')\right] \cdot \left[\bm{\mathcal{M}}^{\Omega}(\vect{r}') - \hat{\mathbf{n}}' \times \vect{E}_{i}^{\Omega,p}(\mathbf{r}')\right]dS, \quad \vect{r} \in V_e.
\end{align}
\end{widetext}

After replacing the external and the internal fields with their corresponding spherical harmonic expansions, then substituting Eq.~\eqref{eq:DyadGreenExpSH_b} for the dyadic Green function $\hat{\mathbf{G}}_{e}^{\Omega}$ valid for the points in the outer region, and finally collecting the Fourier coefficients multiplying radiating spherical harmonics $\vect{M}_{\nu}^{(3)}$ and $\vect{N}_{\nu}^{(3)}$ of the same multi-index $\nu$, we are left with the following matrix equation:
\begin{equation}
	 \mathbf{f}^{\Omega}+ \mathbf{g}^{\Omega(1)} = \mathbf{Q}^{\Omega(1,1)}\mathbf{h}^{\Omega}.
	\label{eq:Q11matrixSH}
\end{equation}
Here, the vector $\mathbf{f}^{\Omega}$ contains the unknown coefficients of the SH external fields, $\{b_{\nu}^{\Omega},a_{\nu}^{\Omega}\}$, and $\mathbf{g}^{\Omega(1)}$ is the source vector with components expressed as surface integrals of products involving tangential components of the nonlinear current sources and regular VSWFs.

Finally, we eliminate the vector $\mathbf{h}^{\Omega}$ between Eqs.~\eqref{eq:Q13matrixSH} and \eqref{eq:Q11matrixSH}, and so arrive at the following matrix equation relating the vector containing the expansion coefficients of the external (scattering) field, $\mathbf{f}^{\Omega}$, and the vectors $\mathbf{g}^{\Omega(3)}$ and $\mathbf{g}^{\Omega(1)}$ holding the expansion coefficients of the nonlinear excitation sources:
\begin{equation}
	 \mathbf{f}^{\Omega} = -\mathbf{g}^{\Omega(1)} + \mathbf{Q}^{\Omega(1,1)}[\mathbf{Q}^{\Omega(1,3)}]^{-1}\mathbf{g}^{\Omega(3)}.
	\label{eq:TmatrixSH}
\end{equation}
This formula can be used to define the $\mathcal{T}$-matrix of the particle that characterizes the nonlinear scattering wave interaction at the SH:
\begin{equation}
	\mathbf{T}^\Omega = \mathbf{Q}^{\Omega(1,1)}[\mathbf{Q}^{\Omega(1,3)}]^{-1}.
	\label{eq:TmatrixfromQSH}
\end{equation}

Similarly to the TMM formulation at the FF, the SH TMM developed here can in principle be used to analyze nonlinear wave scattering from particles with arbitrary morphologies. One important idea related to the nonlinear field calculations, which we want to highlight here, is that in this two-step procedure the accurate evaluation of the linear near-field on the inner side of the particle surface is of paramount importance, as these linear near-fields fully determine the polarization sources at the SH. This aspect of the TMM should be considered carefully, especially when studying metallic particles or particles with surface that deviates significantly from a sphere, as it is well known that in these cases the TMM calculation of the linear near-fields converges slowly.

\section{Formalism extension to multiparticle systems}\label{CLUST}
In this section, we generalize the formalism introduced in the previous section, valid for a single particle, to an arbitrary distribution of particles. This extension is readily achieved by dint of translation-addition matrices \cite{Cruzan,Mackowski,Stout1}, which are used to capture the multiple wave scattering nature of the physical processes investigated.

\subsection{Linear wave scattering}
\begin{figure}[htbp]
	\centering
	\includegraphics[width=\columnwidth]{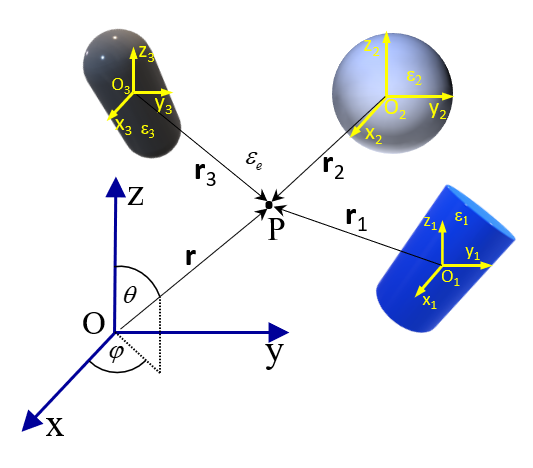}
	\caption{Schematic illustration of a distribution of particles, the global coordinate system with origin at $\mathcal{O}$, and local coordinate systems attached to each of the particles.}
	\label{fig:Mul_Par}
\end{figure}

The physical system under investigation is schematically depicted in Fig.~\ref{fig:Mul_Par}. It consists of a general distribution of $N$ particles whose shape and material parameters are given but arbitrary otherwise. We associate to each particle a (local) coordinate system with origin at $\mathcal{O}_{n}, n=1,\ldots,N$, and describe the entire cluster of particles in a (global) coordinate system with origin at point $\mathcal{O}$, chosen in such a way that the particles are arranged around it in some loosely defined balanced way.

We can express the field exciting the \textit{n}th particle, $n=1,\ldots,N$, as the sum of the incident field at the location of the particle and the fields scattered from all the other particles (after possibly being scattered multiple times among the particles):
\begin{align}\label{eq:IncFielnthPar}
\vect{E}_{n,ex}^{\omega}&(\mathbf{r})= \left[\vect{M}_{\nu}^{(1)}(k_e^\omega \mathbf{r}),\vect{N}_{\nu}^{(1)}(k_e^\omega \mathbf{r})\right]\mathbf{g}^{\omega}\\
&+\sum\limits_{\substack{k=1\\k \ne n}}^{N}\left[\vect{M}_{\nu}^{(3)}(k_e^\omega
\mathbf{r}_k),\vect{N}_{\nu}^{(3)}(k_e^\omega \mathbf{r}_k)\right]\mathbf{f}_k^{\omega}, \quad
n=1,\ldots,N.\nonumber
\end{align}
To simplify the notation, in this equation and in what follows the summation over the multi-index $\nu$ is omitted. This field can be expanded in the local coordinate system associated to the \textit{n}th particle as:
\begin{align}\label{eq:IncFielnthParOrig}
\vect{E}&_{n,ex}^{\omega}(\mathbf{r}_n) = \left[\vect{M}_{\nu}^{(1)}(k_e^\omega \mathbf{r}_n),\vect{N}_{\nu}^{(1)}(k_e^\omega \mathbf{r}_n)\right]\bm{\beta}_{n,0}^{\omega}\mathbf{g}^{\omega}\\
&+\sum\limits_{\substack{k=1\\k \ne n}}^{N}\left[\vect{M}_{\nu}^{(1)}(k_e^\omega
\mathbf{r}_n),\vect{N}_{\nu}^{(1)}(k_e^\omega
\mathbf{r}_n)\right]\bm{\alpha}_{n,k}^{\omega}\mathbf{f}_k^{\omega},~ n=1,\ldots,N.\nonumber
\end{align}
Here, the matrix $\bm{\alpha}_{n,k}^{\omega}$ is the FF irregular translation-addition operator mapping the radiative VSWFs defined in the local coordinate system with origin at $\mathcal{O}_k$ to the regular VSWFs expressed in the local coordinate system with origin at $\mathcal{O}_n$ (see Fig. \ref{fig:Mul_Par}). Similarly, $\bm{\beta}_{n,0}^{\omega}$ is the regular translation-addition matrix computed at the FF $\omega$, which maps the regular VSWFs defined in the global coordinate system with origin at $\mathcal{O}$ to their expansion in the local coordinate system with origin at $\mathcal{O}_n$, associated to the $n$th particle. The total field incident on the $n$th particle, $\vect{E}_{n,ex}^{\omega}(\mathbf{r}_n)$, can itself be expanded into a series of regular VSWFs expressed in the local coordinate system with origin at $\mathcal{O}_n$ and weighted with coefficients stacked up in a vector denoted by $\mathbf{e}_n^{\omega}$. Therefore, we are left with the following equation containing only Fourier expansion coefficients:
\begin{equation}\label{eq:IncFielnthParOrigCoeff}
\mathbf{e}_n^{\omega} = \bm{\beta}_{n,0}^{\omega}\mathbf{g}^{\omega} +\sum\limits_{\substack{k=1\\k
\ne n}}^{N} \bm{\alpha}_{n,k}^{\omega}\mathbf{f}_k^{\omega}, \quad n=1,\ldots,N.
\end{equation}
Note that $\mathbf{e}_n^{\omega}$ describes the field incident onto the $n$th particle of the cluster, so that it plays the r\^{o}le of $\mathbf{g}_n^{\omega}$ for an isolated individual particle labeled by $n$, in the absence of other particles. The key difference between these two vectors is that $\mathbf{e}_n^{\omega}$ consists of both the field directly incident onto the cluster, which reaches the particle $n$ without being scattered by any other particle in the cluster, and the field that reaches the particle $n$ after being scattered (possibly multiple times) by the other particles in the cluster; $\mathbf{g}_n^{\omega}$, on the other hand, consists simply of the incident field exciting the cluster. With this in mind, it should be clear that the relation $\mathbf{f}_n^{\omega}=\mathbf{T}_n^\omega\mathbf{e}_n^{\omega}$ holds, where $\mathbf{T}_n^\omega$ is the $\mathcal{T}$-matrix of the $n$th particle.

Multiplying Eqs.~\eqref{eq:IncFielnthParOrigCoeff} with the $\mathcal{T}$-matrices of individual scatterers $\left\{\mathbf{T}_1^{\omega},\ldots,\mathbf{T}_N^{\omega}\right\}$, respectively, which are precomputed and stored beforehand, we are left with a linear system of equations whose solution gives us the scattering field Fourier coefficients:
\begin{equation}\label{eq:ScattFielnthParOrigCoeff}
	\mathbf{f}_n^{\omega} = \mathbf{T}_n^\omega \bm{\beta}_{n,0}^{\omega}\mathbf{g}^{\omega} +
	\sum\limits_{\substack{k=1\\k \ne n}}^{N} \mathbf{T}_n^\omega
	\bm{\alpha}_{n,k}^{\omega}\mathbf{f}_k^{\omega}, \quad n=1,\ldots,N.
\end{equation}
This linear system of equations can be cast in the following matrix form \cite{Mishchenko0}:
\begin{align}\label{eq:Tmatrixcluster}
	\begin{bmatrix}
		\textbf{I} & -\textbf{T}_{1}^\omega \bm{\alpha}_{1,2}^{\omega} & \cdots &  -\textbf{T}_{1}^\omega \bm{\alpha}_{1,N}^{\omega}\\
		-\textbf{T}_{2}^\omega \bm{\alpha}_{2,1}^{\omega} & \textbf{I} & \cdots &  -\textbf{T}_{2}^\omega \bm{\alpha}_{2,N}^{\omega} \\
		\vdots & \vdots & \ddots & \vdots \\
		-\textbf{T}_{N}^\omega \bm{\alpha}_{N,1}^{\omega} & -\textbf{T}_{N}^\omega
		\bm{\alpha}_{N,2}^{\omega} & \cdots & \textbf{I}
	\end{bmatrix}
	\begin{bmatrix}
		\textbf{f}_{1}^\omega \\
		\textbf{f}_{2}^\omega \\
		\vdots \\
		\textbf{f}_{N}^\omega
	\end{bmatrix}
	=\nonumber\\
	\begin{bmatrix}
		\textbf{T}_{1}^\omega \bm{\beta}_{1,0}^{\omega} \textbf{g}^\omega \\
		\textbf{T}_{2}^\omega \bm{\beta}_{2,0}^{\omega} \textbf{g}^\omega \\
		\vdots \\
		\textbf{T}_{N}^\omega \bm{\beta}_{N,0}^{\omega} \textbf{g}^\omega \\
	\end{bmatrix},
\end{align}
or in the more condensed form,
\begin{equation}
	\mathbf{S}^{\omega} \mathbf{F}^{\omega}  = \mathbf{G}^{\omega}.
	\label{eq:Scattmatclust}
\end{equation}

The matrix $\mathbf{S}^{\omega}$ is the \textit{scattering matrix} of the system of particles. It contains all the information regarding the geometry and material parameters of the cluster. The unknown scattering field coefficients corresponding to all of the particles are collected in the vector $\mathbf{F}^{\omega}$ and are retrieved by solving the linear system \eqref{eq:Scattmatclust}. The corresponding expansion coefficients associated to the fields internal to each of the particles are obtained with the help of Eq.~\eqref{eq:Q11matrixFF}. After the scattering and internal field expansion coefficients are computed for all of the particles the corresponding fields can be directly obtained from Eqs. \eqref{eq:ScatteredVSWFE} and \eqref{eq:InsideVSWFE} (electric fields) and \eqref{eq:ScatteredVSWFH} and \eqref{eq:InsideVSWFH} (magnetic fields). Finally, the scattering cross-section spectrum of the cluster of particles, $\sigma_{sc}^{FF}(\omega)$, is easily obtained using the scattering field coefficients of all of the particles:
\begin{equation}
	\sigma_{sc}^{FF}(\omega) = \frac{1}{(k_e^\omega)^2}\sum\limits_{n,k=1}^{N}
	\mathfrak{Re}({\textbf{f}_{n}^{\omega}}^{\dag}\bm{\beta}_{n,k}^{\omega}\textbf{f}_{k}^{\omega}),
	\label{eq:ScattcrosssectFF}
\end{equation}
where the symbol ``$^{\dag}$'' denotes Hermitian conjugation and $\mathfrak{Re}(z)$ is the real part of the complex number $z$. To obtain this expression one integrates over the scattering angles the differential scattering cross-section of each particle then sums the results over the particles (a detailed derivation of this formula can be found in Ref.~\cite{Stout2}).

\subsection{Wave scattering at the second-harmonic}
The analysis of the wave scattering at the SH follows the same main steps as in the case of the linear wave scattering we just discussed. The main difference between the linear and nonlinear wave scattering amounts to the fact that the sources of the SH field are not located at an infinite distance from the cluster of particles but inside the particles and at their surface. To account for this key difference, we add to the right hand side of the integral equation \eqref{eq:StrattChuSH_a} the excitation field at the SH, representing the source field for the $n$th particle. This excitation field consists of the sum of the fields scattered from all the other particles.

This excitation field at the SH can be expanded into series of regular VSWFs around the origin $\mathcal{O}_n$ associated to the $n$th particle, so that we can establish a relation between the expansion coefficients of the SH internal homogeneous fields corresponding to the $n$th particle and the expansion coefficients of the total nonlinear field impinging onto the particle:
\begin{equation}
	\mathbf{Q}_n^{\Omega(1,3)}\mathbf{h}_n^{\Omega}  = \mathbf{g}_n^{\Omega(3)} - \sum\limits_{\substack{k=1\\k \ne n}}^{N}
	\bm{\alpha}_{n,k}^{\Omega}\mathbf{f}_k^{\Omega},\quad n=1,\ldots,N,
	\label{eq:Q13matrixSH_Clust}
\end{equation}
In these relations, the translation-addition matrices $\bm{\alpha}_{n,k}^{\Omega}$ are evaluated at the SH frequency, $\Omega$. Substituting the internal SH field expansion coefficients $\mathbf{h}_n^{\Omega}$ to Eq.~\eqref{eq:Q11matrixSH} governing the Fourier coefficients of the scattered field at the SH, and iterating through all of the particles, we obtain:
\begin{equation}
	\mathbf{f}^{\Omega}_n = -\mathbf{g}^{\Omega(1)}_n + \mathbf{T}^{\Omega}_n\mathbf{g}^{\Omega(3)}_n -
	\sum\limits_{\substack{k=1\\k \ne n}}^{N} \mathbf{T}_n^\Omega
	\bm{\alpha}_{n,k}^{\Omega}\mathbf{f}_k^{\Omega}, \quad n=1,\ldots,N,
	\label{eq:TmatrixSH_Clust}
\end{equation}
These relations can be expressed in matrix form as:
\begin{align}\label{eq:Tmatrixcluster_SH}
	\begin{bmatrix}
		\textbf{I} & \textbf{T}_{1}^\Omega \bm{\alpha}_{1,2}^{\Omega} & \cdots &  \textbf{T}_{1}^\Omega \bm{\alpha}_{1,N}^{\Omega}\\
		\textbf{T}_{2}^\Omega \bm{\alpha}_{2,1}^{\Omega} & \textbf{I} & \cdots &  \textbf{T}_{2}^\Omega \bm{\alpha}_{2,N}^{\Omega} \\
		\vdots & \vdots & \ddots & \vdots \\
		\textbf{T}_{N}^\Omega \bm{\alpha}_{N,1}^{\Omega} & \textbf{T}_{N}^\Omega
		\bm{\alpha}_{N,2}^{\Omega} & \cdots & \textbf{I}
	\end{bmatrix}
	\begin{bmatrix}
		\textbf{f}_{1}^\Omega \\
		\textbf{f}_{2}^\Omega \\
		\vdots \\
		\textbf{f}_{N}^\Omega
	\end{bmatrix}
	=\nonumber\\
	\begin{bmatrix}
		-\textbf{g}_1^{\Omega(1)}+\textbf{T}_{1}^\Omega \textbf{g}_1^{\Omega(3)} \\
		-\textbf{g}_2^{\Omega(1)}+\textbf{T}_{2}^\Omega \textbf{g}_2^{\Omega(3)} \\
		\vdots \\
		-\textbf{g}_N^{\Omega(1)}+\textbf{T}_{N}^\Omega \textbf{g}_N^{\Omega(3)} \\
	\end{bmatrix},
\end{align}
or, more concisely,
\begin{equation}
	\mathbf{S}^{\Omega} \mathbf{F}^{\Omega}  = \mathbf{G}^{\Omega}.
	\label{eq:Scattmatclust_SH}
\end{equation}

Similarly to the wave scattering at the FF, $\mathbf{S}^{\Omega}$ represents the system scattering matrix at the SH, the vector $\mathbf{F}^{\Omega}$ holds the Fourier coefficients of the scattering field corresponding to all of the particles, and the vector $\mathbf{G}^{\Omega}$ contains the information on the nonlinear polarization sources responsible for the generation of the optical field at the SH. After the SH scattering field expansion coefficients corresponding to all of the particles are computed by solving Eq.~\eqref{eq:Scattmatclust_SH}, for each particle one can compute \textit{via} Eq.~\eqref{eq:Q11matrixSH} the expansion coefficients corresponding to the internal \textit{homogeneous} SH field. The associated scattering and internal homogeneous fields are computed, respectively, from Eqs.~\eqref{eq:ScatteredVSWFE_SH} and \eqref{eq:InsideVSWFE_SH} (electric fields) and Eqs.~\eqref{eq:ScatteredVSWFH_SH} and \eqref{eq:InsideVSWFH_SH} (magnetic fields). Finally, in order to compute the total fields inside each particle, the particular (inhomogeneous) solution defined by Eqs.~\eqref{eq:PartsolSH} must be added to the homogeneous one. This is conveniently done by expanding the product of linear electric fields inside the particle into Fourier series of VSWFs \cite{Varshalovich}. The SH scattering cross-section spectra are then calculated in an analogous fashion to the procedure used at the FF, as per \eqref{eq:ScattcrosssectFF}, with all the physical quantities being evaluated at SH frequency.

Let us now briefly discuss the particular case in which the nanoparticles are located onto a planar substrate. This is an important case from a practical point of view, as the substrate can affect the optical response of the cluster of nanoparticles. The influence of the substrate can be taken into account in two ways. In a first approach, one can use image techniques, whereby for each nanoparticle lying onto the substrate one adds its image located inside it. Then, one can treat using the method developed in this work this new, larger cluster. In an alternative, more rigorous approach, one would replace in our formalism the free-space Green function with the Green function of an optical slab with finite thickness, for which there is a well-known analytical formula \cite{Jackson}.

Before ending this section, we would like to say that we have implemented in a freely available high-performance computational (HPC) code, \textsf{OPTIMET-3D} \cite{optimet}, the numerical method we just described. The computer program is a fast, parallel, and highly scalable \textsf{C++} code, which has been installed and extensively tested on several local HPC clusters, as well as \textsf{ARCHER2}, UK's national HPC platform. The linear systems of equations at the fundamental frequency and second harmonic are solved using routines available in \textsf{ScaLAPACK} and the parallel iterative generalized minimal residual method. Moreover, the coupling submatrices representing interactions between nanoparticles were approximated and compressed \textit{via} the low-rank adaptive cross-approximation algorithm. This speeds up significantly the computations, especially when a large number of nanoparticles is considered. These code optimizations allowed us to consider the linear and nonlinear optical response of clusters of hundreds and even thousands of nanoparticles.

\section{Applications and discussion}\label{NumRes}
In this section we will illustrate how the numerical method developed in the preceding section can be used to compute the scattering cross-section spectra of systems of particles with different morphology, the calculations being performed both at the fundamental and second-harmonic frequencies. In the case in which the particles are spherical, we compare our results with analytical solutions \cite{Forestiere1} (for a single sphere), or predictions of the TMM with analytically computed $\mathcal{T}$-matrices (for a multiple sphere system). Because in this latter case the analytic calculation of the matrix elements is based on the Mie theory \cite{Mie1}, we call it \textit{Mie-TMM} \cite{Sekulic1}. In the case of particles with non-spherical shape, we validate our results against results obtained using \textsf{CST Studio}\textsuperscript{\textregistered} \cite{CSTStudio}, a commercial software based on the FEM. We consider particles made of two centrosymmetric optical materials widely used in nanophotonics, namely silicon (Si) and gold (Au). Our choice was guided by the fact that both these optical materials present challenging testing conditions for any numerical method, namely silicon particles have large electric permittivity and hence scatter significantly the incident field,  whereas (plasmonic) particles made of Au possess plasmon resonances that induce strongly inhomogeneous, enhanced near-fields.

The cluster of particles is assumed to be embedded in vacuum, whereas the excitation field is a $\hat{\bm{\theta}}$ polarized plane wave whose propagation direction is defined by the incident angles $\theta_{inc} = \pi/4$ and $\phi_{inc} = \pi/2$. The frequency-dependent permittivity of silicon and gold is interpolated from the experimental data provided in Refs.~\cite{Schinke} and \cite{Johnson}, respectively. The tensor components of the nonlinear surface susceptibility $\bm{\hat{\chi}_{s}^{(2)}}$ of gold, as well as its bulk nonlinear susceptibility $\gamma$, are calculated using the free-electron hydrodynamic model \cite{Sipe1}:
\begin{subequations}\label{eq:RS}
	\begin{align}
		&\chi_{s,\perp\perp\perp}^{(2)} = -\frac{a}{4} \left[\epsilon_{r}(\omega) - 1\right]
		\frac{e}{m\omega^2},\label{eq:ksiperperper}\\
		&\chi_{s,\parallel\perp\parallel}^{(2)} =\chi_{s,\parallel\parallel\perp}^{(2)} = -\frac{b}{2} \left[\epsilon_{r}(\omega) -1\right]
		\frac{e}{m\omega^2}, \label{eq:ksiparparper} \\
		&\gamma  = -\frac{d}{8} \left[\epsilon_{r}(\omega) - 1\right] \frac{e}{m\omega^2},
		\label{eq:gammaBul}
	\end{align}
\end{subequations}
where the Rudnick-Stern coefficients \cite{Rudnick} are $a=1$, $b=-1$, and $c=1$ \cite{Sipe1}, whereas $e$ and $m$ denote the charge and the mass of the electron, respectively. The corresponding nonlinear susceptibilities of silicon are assumed to be constant in the whole frequency range of interest, with values equal to $\chi_{s,\perp\perp\perp}^{(2)} =\SI{65e-19}{\meter\squared\per\volt}$,
$\chi_{s,\parallel\perp\parallel}^{(2)} =\chi_{s,\parallel\parallel\perp}^{(2)} = \SI{3.5e-19}{\meter\squared\per\volt}$, and $\gamma =
\SI{1.3e-19}{\meter\squared\per\volt}$ \cite{Falasconi}. For both optical materials, gold and silicon, we set $\chi_{s,\perp\parallel\parallel}^{(2)}=0$, which is a commonly used assumption \cite{Corvi},\cite{Guyot}.

\subsection{Convergence tests}
\begin{figure}[!b]
	\centering
	\includegraphics[width=\columnwidth]{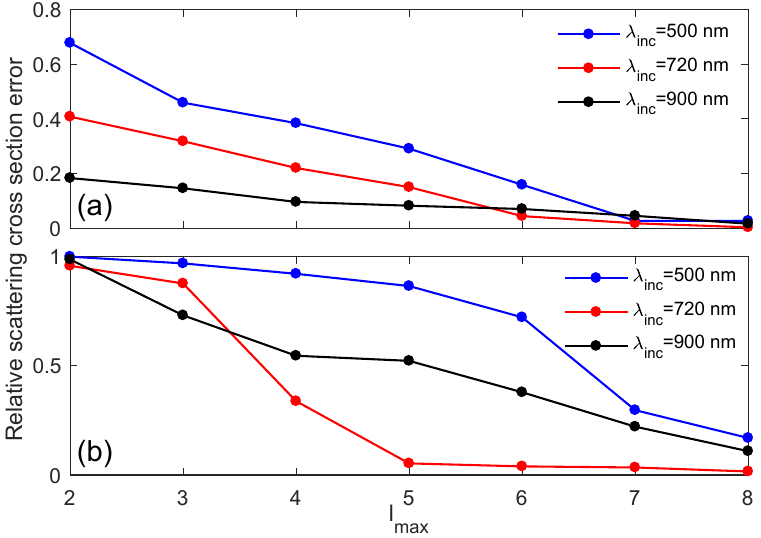}
	\caption{Relative scattering cross-section error of two Si spheres at (a) FF and (b) SH.}
	\label{fig:conv_trends_sph}
\end{figure}
We begin our investigation of the properties of the numerical method developed in this study with an analysis of its convergence characteristics. For this purpose, we choose two representative numerical examples and monitor the evolution of the relative scattering cross-section error with the maximum expansion order, $l_{max}$. This analysis was performed for several values of the frequency of the incident wave. In the first example we considered two silicon nanospheres with radii $R_1=\SI{250}{\nano\meter}$ and $R_2=\SI{200}{\nano\meter}$ centered at $O_1(0,\,0,\,0)$ and $O_2(0,\,0,\,\SI{550}{\nm})$. In the second example the particle system consists of two spheroids, a prolate spheroid with equatorial and polar radii $a_1=\SI{200}{\nano\meter}$ and $c_1=\SI{300}{\nano\meter}$, respectively, and an oblate spheroid with radii $a_2=\SI{320}{\nano\meter}$ and $c_2=\SI{230}{\nano\meter}$. The two spheroids are made of gold and centered at $O_1(0,\, \SI{-350}{\nm},\,0)$ and $O_2(0,\,\SI{350}{\nm},\,0)$, respectively.
\begin{figure}[!t]
	\centering
	\includegraphics[width=\columnwidth]{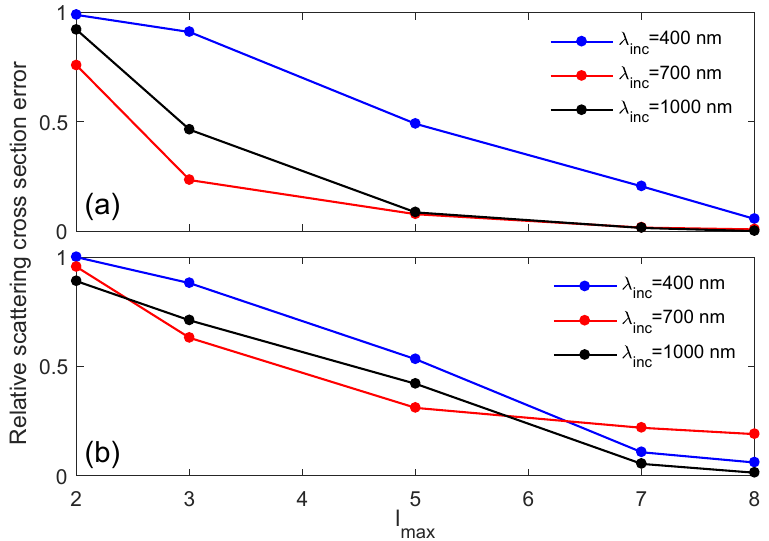}
	\caption{Relative scattering cross-section error of two Au spheroids at (a) FF and (b) SH.}
	\label{fig:conv_trends_ell}
\end{figure}

We define the relative scattering cross-section error as:
\begin{align}\label{eq:norm}
\mathcal{E}^{\omega/\Omega} =
\left|\frac{\sigma_{l_{max}}^{\omega/\Omega}-\sigma_{REF}^{\omega/\Omega}}{\sigma_{REF}^{\omega/\Omega}}\right|.
\end{align}
The reference scattering cross-section, $\sigma_{REF}^{\omega/\Omega}$, is calculated within the Mie-TMM formalism \cite{Sekulic1} for nanospheres, and, for the case of spheroids, with the TMM method introduced in this work and expansion order $l_{max}=10$. According to Figs.~\ref{fig:conv_trends_sph} and \ref{fig:conv_trends_ell}, a reduction in the relative scattering cross-section error, both in FF and SH regimes, is observed as the number of spherical harmonics included in the Fourier expansion increases. This is the usual trend observed when a numerical method converges. We also note a general trend, more evident in the case of scattering cross-section error evaluated at the FF, namely that the error is smaller for larger values of the wavelength of the incident wave. This is explained by the fact that as the wavelength increases the wave scattering process approaches the quasi-static regime described by the first order of the series expansion. This conclusion is not necessarily valid at the SH, as in this case the nonlinear source distribution can be markedly affected by the excitation of optical resonances of the particles, a phenomenon that is strongly dependent on the wavelength.

It should be noted that these convergence tests can be used to infer the values of critical parameters, such as the maximum expansion order, $l_{max}$, and the number of triangles used to produce the meshes covering the particles, that lead to accurate results. Importantly, we considered the most challenging cases, namely dielectric and metallic particles with large electric permittivity. This means that modelling particle systems with smaller refractive index contrast, such as colloidal systems, would require smaller computational resources.

\subsection{Wave scattering from nanospheres}
In this numerical example, we choose a nanosphere made of either gold or silicon and of radius equal to $R=\SI{50}{\nano\meter}$ in both cases. Nanospheres of this characteristic size are commonly used in different applications, including plasmonic sensing and colloidal chemistry. Although it is a simple physical system, it has the advantage that analytic solutions to the linear and nonlinear scattering problems exist, so that it provides an effective validation test of our numerical method. Thus, for this system we calculated the spectra of the scattering cross-section, at both the FF and SH, in a spectral domain of practical interest. We meshed the surface of each nanosphere into 2,000 triangles, which was enough to reach convergence. The scattering cross-section spectra were obtained with the method introduced in this study, where the corresponding surface integrals were computed numerically over the boundary tessellation, and the results were compared to the conclusions derived from the analytical Mie theory \cite{Forestiere1}. In both approaches, numerical and analytical, we used a maximum expansion order of $l_{max}=6$, which corresponds to a total of 96 VSWFs per sphere at the FF and SH.
\begin{figure}[!b]
	\centering
	\includegraphics[width=\columnwidth]{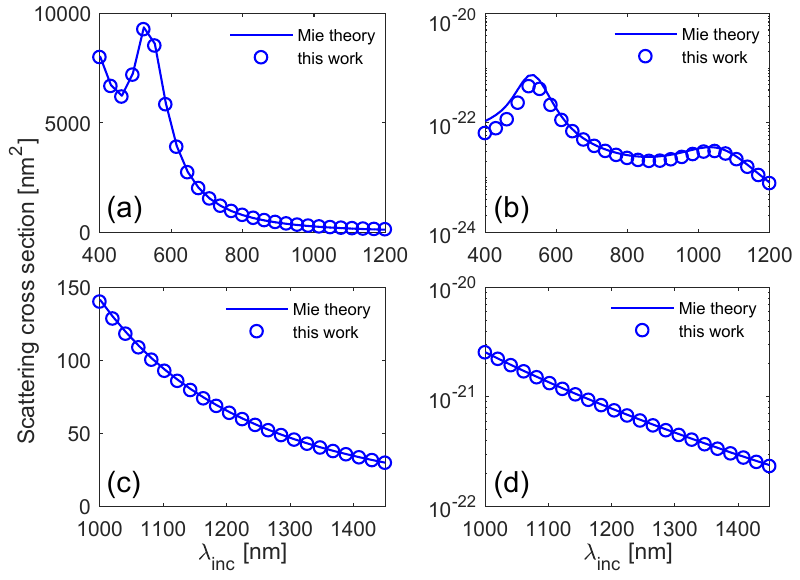}
	\caption{Scattering cross-sections at FF (left panels) and SH (right panels) for a gold (top panels) and silicon (bottom panels) sphere with radius $R=\SI{50}{\nano\meter}$.}
	\label{fig:SCS_singsph}
\end{figure}

The spectra computed using this procedure are plotted in Fig.~\ref{fig:SCS_singsph}. It is interesting to notice that the FF and SH scattering spectra of the gold particle exhibit a narrow spectral peak at the incident wavelength $\lambda_{inc}=\SI{520}{\nano\meter}$, as per Figs.~\ref{fig:SCS_singsph}(a) and \ref{fig:SCS_singsph}(b). These peaks can be attributed to the plasmonic resonances of a gold nanosphere. On the other hand, Mie resonances, characteristic to dielectric particles, are not observed for the silicon nanosphere in the frequency range considered here, as for particles with size considered in this example they exist for shorter wavelengths. Furthermore, according to the data presented in Fig.~\ref{fig:SCS_singsph}, there is a good agreement, both at fundamental and second-harmonic frequencies, between scattering cross-section spectra obtained with the analytical and numerical procedures.
\begin{figure}[!t]
	\centering
	\includegraphics[width=\columnwidth]{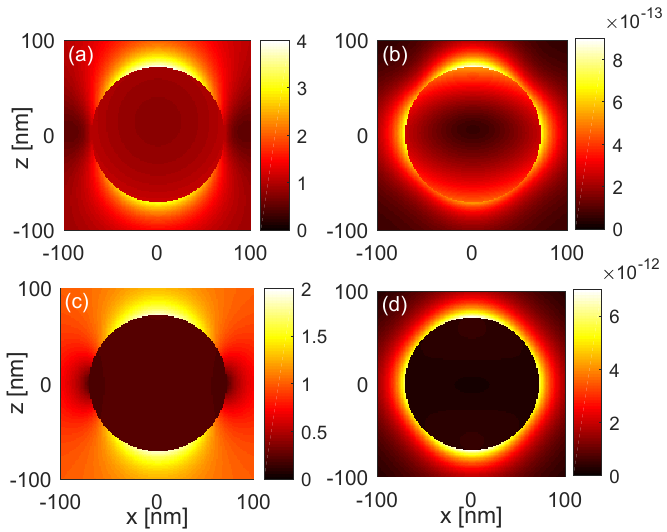}
	\caption{Electric field amplitude in the $y=0$ median plane of a Au (top panels) and Si (bottom panels) sphere with radius of \SI{50}{\nano\meter}, at FF (left panels) and SH (right panels).}
	\label{fig:Field_singsph}
\end{figure}

In Fig.~\ref{fig:Field_singsph} we show the spatial distribution of the amplitude of the electric near-field for the silicon and gold nanospheres, computed at the FF and SH. We choose the incident wavelength $\lambda_{inc}=\SI{520}{\nano\meter}$ for the gold particle, so as to excite the plasmonic resonance seen in the spectra of the scattering cross-section corresponding to the FF and SH. In the case of the silicon nanosphere we pick an off-resonance incident wavelength $\lambda_{inc}=\SI{1100}{\nano\meter}$ for our field plots, since Mie resonances are not present in the frequency range considered here.

There are several ideas illustrated by these field profiles. First, the near-field distribution at the FF shows the dipolar character of the scattering process at the wavelength considered. At the SH, on the other hand, the near-fields in the two cases are of a different nature, namely the field in the case of the Au nanosphere is quadrupolar, whereas for the Si nanosphere it is almost radially symmetric. Second, in both cases the SH near-field is confined primarily at the surface of the nanosphere, suggesting that the contribution of the nonlinear surface polarization dominates when compared to that of the bulk polarization. Moreover, if only the contributions of the bulk SHG are compared, it appears that it is significantly larger in the case of the Au nanosphere. This conclusion can be inferred from the spatial distribution of the electric field inside the particle at the FF, as the gradients of its components determine the nonlinear bulk polarization. Thus, since the wavelength is much larger than the size of the sphere, the quasistatic regime is valid and hence one can view the particle as if it were placed in a uniform electric field. Consequently, the electric field inside is practically uniform and the bulk nonlinear polarization vanishes.
\begin{figure}[!b]
	\centering
	\includegraphics[width=\columnwidth]{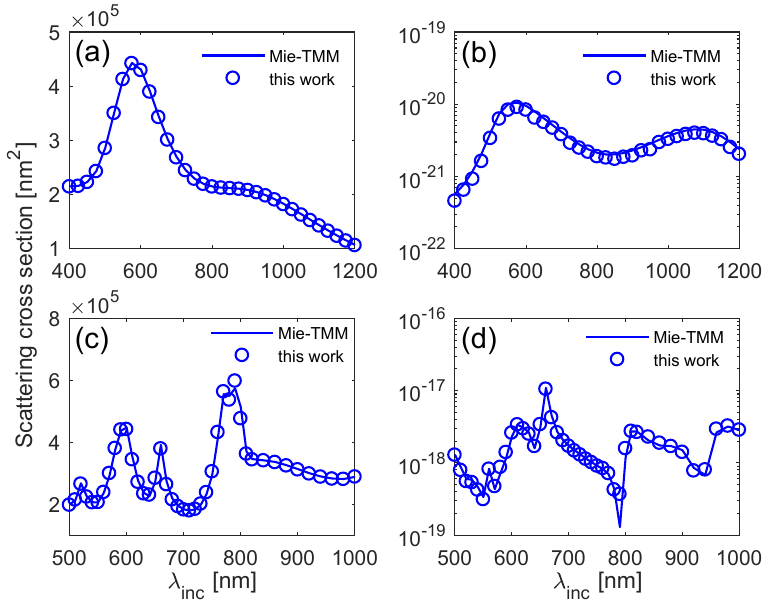}
	\caption{Scattering cross-section spectra at the FF (left panels) and SH (right panels) for three nanospheres made of Au (top panels) and Si (bottom panels).}
	\label{fig:SCS_threesph}
\end{figure}

As part of the second validation test, we analyze the linear and nonlinear wave scattering from three nanospheres made of gold or silicon, with centers located at $O_1$(\SI{-100}{\nm}, \SI{-100}{\nm}, \SI{100}{\nm}), $O_2$(\SI{-150}{\nm}, \SI{150}{\nm}, \SI{100}{\nm}), and $O_3$(\SI{200}{\nm}, \SI{200}{\nm}, \SI{100}{\nm}), and corresponding radii equal to $R_1=\SI{50}{\nm}$, $R_2=\SI{100}{\nm}$, and $R_3=\SI{150}{\nm}$. Following the numerical scheme presented in this work, we meshed each sphere with 2000 triangles and expanded the fields associated to each sphere into 160 VSWFs. As a result, we have used a total of 480 harmonics for both the FF and SH frequency. With this surface discretization and field expansions we then computed the surface integrals defining the $\mathcal{T}$-matrices at the FF and SH and subsequently evaluated the spectra of the corresponding scattering cross-sections. Alternatively, we calculated analytically the elements of the $\mathcal{T}$-matrices, which is possible in the case of spherical particles, and then the spectra of the scattering cross-sections. In other words, we employed what we call the Mie-TMM approach \cite{Sekulic1}. The results obtained using the two methods are compared in Fig.~\ref{fig:SCS_threesph}, where an excellent agreement between the predictions of the method developed here and analytical calculations can be observed both at the FF and SH.

\subsection{Wave scattering from prolate spheroids}
We now consider more challenging examples, which are not amenable to an analytical treatment. The first such numerical example consists of linear and nonlinear electromagnetic wave scattering from prolate spheroids made of either gold or silicon and with different aspect ratio, $\xi=c/a$, where $c$ and $a$ are the polar and equatorial radii of a spheroid, respectively. In this numerical experiment, we keep the equatorial radius constant, $a=\SI{50}{\nano\meter}$, and vary the polar radius from $c_{min}=\SI{50}{\nano\meter}$ (sphere) to $c_{max}=\SI{100}{\nano\meter}$. The numerical results for high aspect ratios, $\xi>2$, are not presented because of the slow convergence, especially in the case of SH frequency. This problematic behavior arises from the fact that the electric fields at the FF evaluated on the surface of the particles converge very slowly as the number of spherical harmonics increases \cite{Somerville}. All the spheroids under consideration are meshed in approximately 2800 triangles and analyzed with 70 vector spherical harmonics each, both at the FF and SH frequency. To give a better sense vis-\`{a}-vis the resources required by our method, the PMCHWT integral formulation employing discretization with RWG-MoM on the same computational mesh would require just for solving the linear scattering problem around 4200 basis functions per equivalent current densities (electric and magnetic) totalling 8400 unknowns.
\begin{figure}[!b]
	\centering
	\includegraphics[width=\columnwidth]{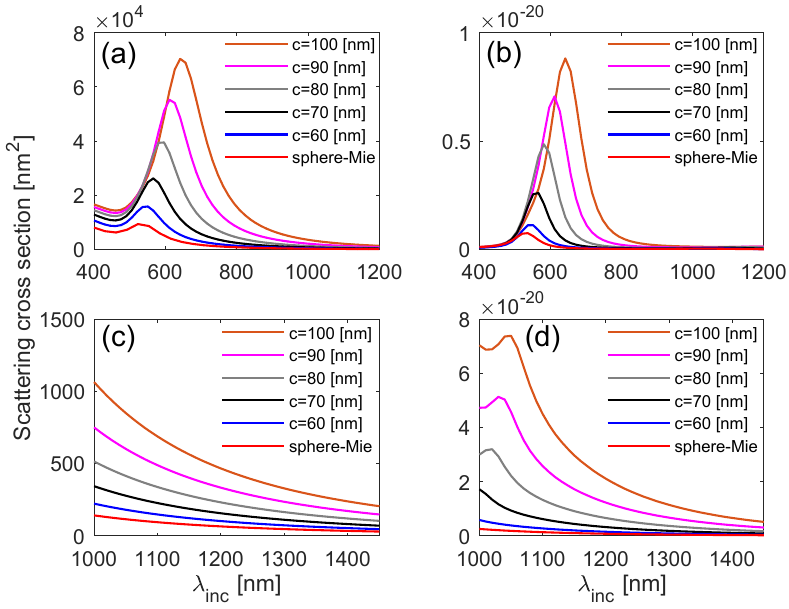}
	\caption{Scattering cross-sections at FF (left panels) and SH (right panels) for prolate spheroids made of Au (top panels) and Si (bottom panels), computed for $a=\SI{50}{\nano\meter}$ and different values of the aspect ratio $\xi=c/a$.}
	\label{fig:SCS_singellip}
\end{figure}

The spectra of the scattering cross-sections corresponding to these spheroids are presented in Fig.~\ref{fig:SCS_singellip}. It can be seen in this figure that, as the aspect ratio $\xi\to1$, both at the FF and SH the scattering cross-section spectra of the spheroids converge to that of a sphere, depicted with red lines in Fig.~\ref{fig:SCS_singellip} and computed analytically. This is the expected evolution because as $\xi\to1$ the spheroids become less elongated and approach a spherical shape. Moreover, the scattering cross-sections increase as the radius $c$ increases, which is explained by the fact that the geometrical area of the spheroids increases, too. Importantly, it can be seen in Figs.~\ref{fig:SCS_singellip}(a) and \ref{fig:SCS_singellip}(b) that the wavelengths of the corresponding plasmonic resonances are redshifted as $\xi$ increases, which is a well-known size-dispersion property of plasmonic resonance wavelength of nanoparticles.
\begin{figure}[!t]
	\centering
	\includegraphics[width=\columnwidth]{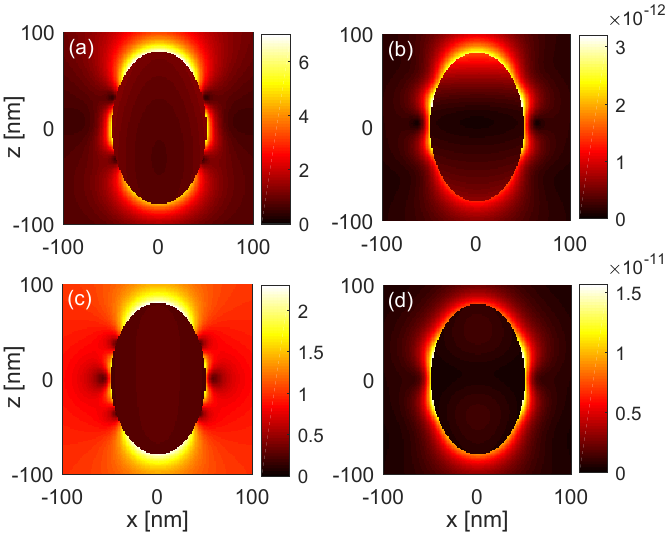}
	\caption{Spatial distribution of the electric field amplitude evaluated in the $y=0$ plane passing through the middle of a prolate spheroid with radii $a=\SI{50}{\nano\meter}$ and $c=\SI{80}{\nano\meter}$, at the FF (left panels) and SH (right panels) for Au (top panels) and Si (bottom panels).}
	\label{fig:Field_singellip}
\end{figure}

In Fig.~\ref{fig:Field_singellip} we plot the spatial profile of the electric near-field amplitude, $\lvert\vect{E}\rvert$, computed in the $y=0$ plane passing through the middle of a prolate spheroid with radii $a=\SI{50}{\nano\meter}$ and $c=\SI{80}{\nano\meter}$. The spheroid is centered at the origin of the coordinate system and the plots are evaluated at the incident wavelengths $\lambda_{inc}=\SI{580}{\nano\meter}$ and $\lambda_{inc}=\SI{1020}{\nano\meter}$ for the gold and silicon spheroids, respectively. These wavelengths correspond to the peaks in the SH scattering cross-section spectra indicating the excitation of plasmonic and Mie resonances, respectively. The most important conclusion revealed by these field profiles is that, as in the case of a sphere, most of the SH field is generated at the surface of the nanoparticles.

\subsection{Wave scattering from a dimer of spheroids}
The next particle system we considered consisted of two identical prolate spheroids with radii $a=\SI{50}{\nm}$ and $c=\SI{70}{\nm}$ located at $O_1(0,\,0,\,0)$ and $O_2(0,\,0,\,\SI{200}{\nm})$. Thus, the spheroids are separated by a gap with size $g=\SI{60}{\nm}$, the center-to-center distance being $d=\SI{200}{\nm}$. We considered only the case of spheroids made of gold, because in this case the presence of the narrow gap between the spheroids can lead to interesting phenomena, such as the formation of hot-spots inside the gap between the two spheroids. Note that such hot spots can also form at the two ends of the two spheroids. As in the previous cases, we computed the scattering cross-section spectra and spatial field profiles both at the FF and SH.
\begin{figure}[!t]
	\centering
	\includegraphics[width=\columnwidth]{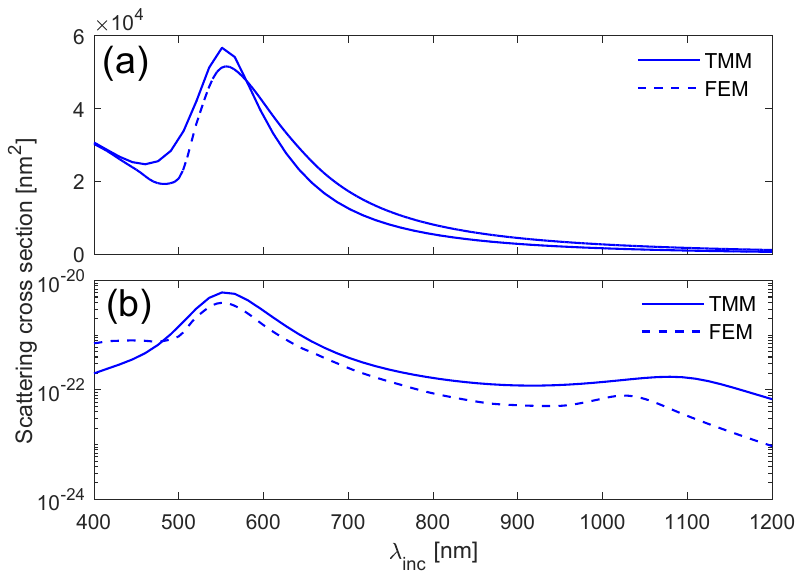}
	\caption{Scattering cross-section at (a) the FF and (b) SH for two prolate spheroids made of Au.} \label{fig:SCS_twoellip}
\end{figure}

The scattering cross-section spectra are plotted in Fig.~\ref{fig:SCS_twoellip}, for the fundamental and second-harmonic frequencies. We calculated these spectra using the TMM presented in this study and compared the results to the spectra obtained using \textsf{CST Studio}\textsuperscript{\textregistered}. We meshed each spheroid in 2,800 triangles and both in the FF and SH regimes we decomposed the electromagnetic fields by employing a total of 140 spherical harmonics (\textit{i.e.}, 70 harmonics per spheroid). This is a substantial reduction of unknown quantities as compared to the approach based on the FEM, in which case about 67,000 elements were necessary. Figure~\ref{fig:SCS_twoellip} illustrates a good agreement between the results obtained by the two methods, both at the FF and SH and in the entire frequency range considered in our simulations. Despite this rather good agreement, it can be seen in Fig.~\ref{fig:SCS_twoellip} that there are some differences between the predictions of the two methods. However, it should be noted that the scattering cross-sections can be viewed as physical quantities that describe the overall optical response of the cluster, and therefore the corresponding error in the computation of these quantities represents a cumulative error given by the summation of the error in the differential scattering cross-section over all possible directions. In other words, the error in the computation of the scattering cross-section is a maximal error characterizing our numerical method.

We stress that this large difference between the number of unknowns needed to represent the electromagnetic fields in the two methods, makes the TMM the clearly preferred choice when one is interested in the optical response of clusters containing a large number of particles. In particular, whereas the use of the FEM becomes problematic even for a few particles, the TMM can be used to investigate clusters with hundreds or even thousands of particles.

One interesting phenomenon can be observed when one compares the spectra in Figs.~\ref{fig:SCS_singellip}(b) and \ref{fig:SCS_twoellip}(b), which correspond to the SHG spectra of a single and dimer of prolate spheroids made of Au, respectively. More specifically, the spectrum of the single particle shows a single resonance, whereas the dimer spectra have two resonances, one at roughly the wavelength of the single-particle resonance and another one at almost twice this wavelength. The latter resonance corresponds to the overall optical response of the ensemble of the two particles, whereas the spectral shift of the former is due to inter-particle interactions.

\begin{figure}[!t]
	\centering
	\includegraphics[width=\columnwidth]{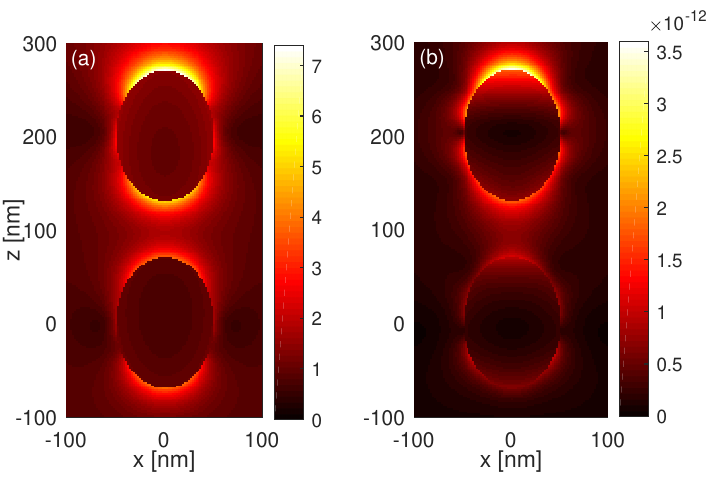}
	\caption{Electric field amplitude at FF (a) and SH (b) computed in the $y=0$ plane passing through the middle of two identical prolate Au spheroids with radii $a=\SI{50}{\nano\meter}$ and $c=\SI{70}{\nano\meter}$ located on the \textit{z}-axis.}
	\label{fig:Field_twoellip}
\end{figure}

In Figs.~\ref{fig:Field_twoellip}(a) and \ref{fig:Field_twoellip}(b) we present the spatial profile of the electric field magnitude at the FF and SH, respectively, computed in the $y=0$ plane passing through the middle of the dimer of spheroids. These field profiles correspond to an incident wavelength $\lambda_{inc}=\SI{550}{\nano\meter}$, for which plasmonic resonances can be observed in the spectra of the scattering cross-sections at the FF and SH. Since the dimer is not symmetrically located with respect to the origin of the coordinate system, the spatial profiles of both the linear and nonlinear fields exhibit a certain asymmetry. Moreover, there is a significant optical near-field in-between the particles, which facilitates their optical coupling. Importantly, one can observe in these field profiles the hot spots created at the sharper ends of the spheroids. The hot spots at the FF are due to the well-known effect of strong field enhancement near the edges and corners of metallic particles, whereas the ones at the SH are due to the enhanced nonlinear polarization induced by the strong surface field associated to the hot spots at the FF.

\subsection{Wave scattering analysis of three particles}
As the last example that illustrates some powerful features of our numerical method, we consider the linear and nonlinear wave scattering from a cluster of three particles with different morphologies. We assume that the particle cluster comprises an oblate spheroid with radii $a_1=\SI{50}{\nm}$ and $c_1=\SI{30}{\nm}$ centered at $O_1(0,\,-\SI{100}{\nm},\,\SI{100}{\nm})$, a prolate spheroid with radii $a_2=\SI{50}{\nm}$ and $c_2=\SI{70}{\nm}$ and center at $O_2(0,\,\SI{100}{\nm},\,\SI{100}{\nm})$, and a sphere with radius $r=\SI{50}{\nm}$ and located at $O_3(\SI{150}{\nm},\,\SI{100}{\nm},\,\SI{100}{\nm})$. The spheroids are oriented such that the \textit{z}-axis is parallel to their polar axis. All the particles are assumed to be made of gold, so that the resonant nature of the scattering process would be more apparent.

The scattering cross-section results used to benchmark our method are computed with \textsf{CST Studio}\textsuperscript{\textregistered} using a finite-element mesh containing 42,400 tetrahedra. In our calculations based on the TMM developed here, each nanoparticle is meshed using about 2,500 triangles and analyzed with 96 spherical harmonics, hence leading to a total of 288 degrees of freedom for the FF and SH computations. This again represents a substantial reduction of memory requirements as compared to the resources needed in the FEM computations.
\begin{figure}[!b]
	\centering
	\includegraphics[width=\columnwidth]{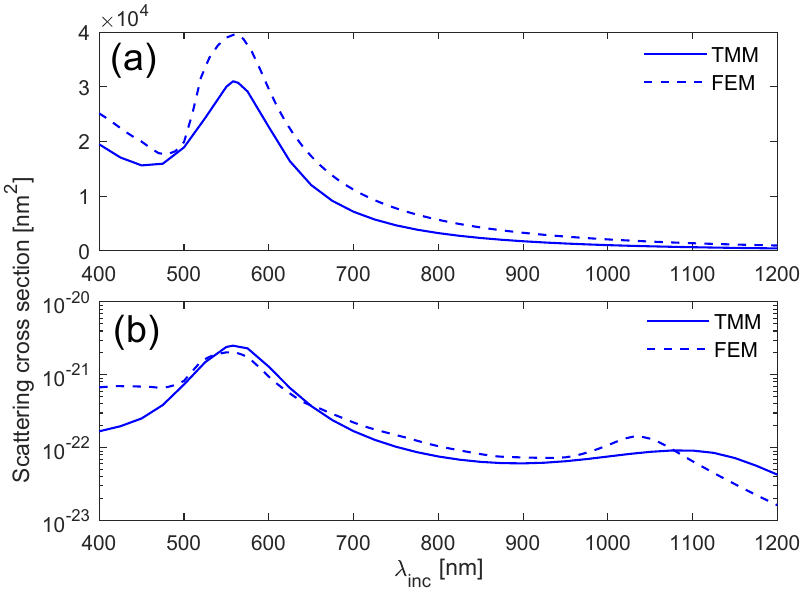}
	\caption{Scattering cross sections at (a) FF and (b) SH for three Au particles with different shapes.}
	\label{fig:SCS_threearb}
\end{figure}

In view of Fig.~\ref{fig:SCS_threearb}, where the data obtained using the two numerical methods are summarized, there is a good agreement both in the linear and nonlinear regimes between the FEM and TMM predictions, in the entire frequency range. As it transpires from these plots, the differences between the results obtained with the two methods are more noticeable near the plasmonic resonances. This is explained by the slow convergence of the near-field, which is the main physical quantity that determines the nonlinear optical field. Whereas the accuracy of the FEM results could have been improved by further increasing the number of finite-elements used in the computations, we were precluded to do this given the extent of the computational resources available to us.

\section{Application to nanoparticle clusters possessing dual optical resonances}\label{ApplBIC} %
\begin{figure}[!b]
	\centering
	\includegraphics[width=\columnwidth]{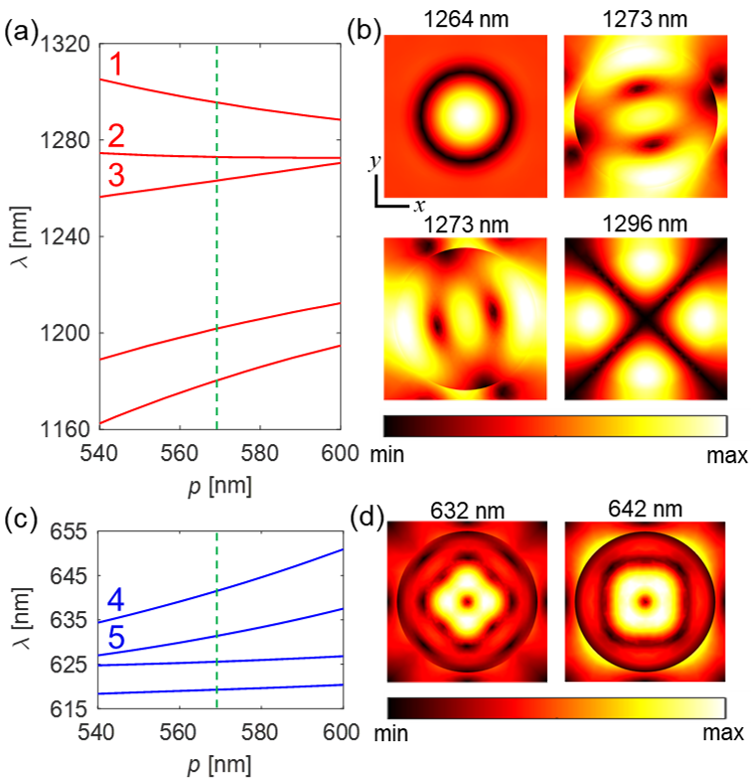}
	\caption{(a) Resonance wavelength vs. period, $p$, determined for an infinite nanosphere metasurface at the $\Gamma$-point. (b) Field profiles of the BIC modes, determined for $p=\SI{570}{\nm}$. (c), (d) The same as in (a) and (b), respectively, but calculated in the spectral region where SH is generated. The vertical green dashed lines indicate the period $p=\SI{570}{\nm}$. The numbers \emph{1} through \emph{5} label the BIC bands.}
	\label{fig:BSMP}
\end{figure}
After investigating the convergence properties of our numerical method and how it performs in several well-known particle configurations, in this section we illustrate its power and versatility by applying it to a particle configuration that is practically intractable with currently available numerical methods. To this end, we proceeded to design an optical cluster for SHG enhancement, by choosing its geometrical parameters such that it possesses optical resonances both at the FF and SH. To achieve this dual-resonance property in a particle cluster that is relevant from an experimental point of view, we considered a planar distribution of 177 silicon nanospheres with radius $R=\SI{250}{\nm}$ placed at the vertices of a square lattice lying in the $(x,y)$-plane. Specifically, the cluster was constructed by placing a nanosphere at the origin of the lattice, then adding 4 nanospheres at the location of the nearest neighbours, then placing 4 nanospheres at the location of the next-nearest-neighbours, and so on, until all 177 spheres were placed in the cluster.
\begin{figure}[!b]
	\centering
	\includegraphics[width=\columnwidth]{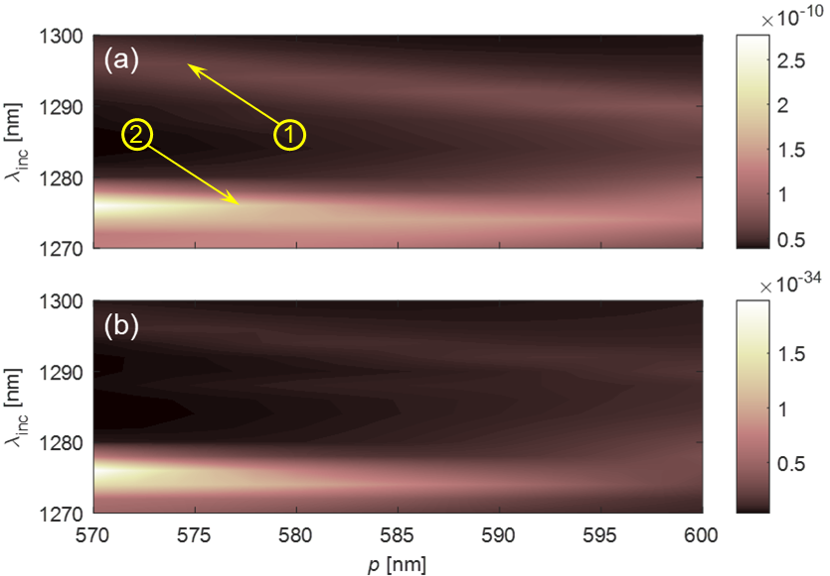}
	\caption{(a), (b) Dispersion maps of the scattering cross-section at the FF and SH, respectively. The cluster consists of 177 silicon nanospheres with radius $R=\SI{250}{\nm}$ arranged in a square array, with the distance between adjacent nanospheres being $p$.}
	\label{fig:scs_cl}
\end{figure}

To investigate the optical properties of this nanosphere optical cluster, we first considered an infinite periodic distribution of silicon nanospheres. For this metasurface, we computed the photonic band structure and the spatial field distribution of certain optical modes. In particular, using \textsf{COMSOL Multiphysics}\textsuperscript{\textregistered} \cite{COMSOL}, a commercially available software, we have computed the resonance wavelength of the optical modes corresponding to the $\Gamma$-point ($k_{x}=k_{y}=0$) for different values of the metasurface period, $p$. We performed these computations in the spectral domains where the FF optical beam is injected and the SH is generated. Amongst these modes, there are some with nontrivial topological properties, the so-called bound-states in the continuum (BICs) \cite{Hsu,Wang}. They are characterized by a nonzero topological charge, which is defined as:
\begin{equation}\label{eq:topch}
	q = \frac{1}{2\pi}\oint_{\mathcal{C}}d\mathbf{k}\cdot\nabla_{\mathbf{k}}\phi(\mathbf{k}),
\end{equation}
where $\phi(\mathbf{k})$ is the angle of the polarization vector and $\mathcal{C}$ is a closed simple path in $\mathbf{k}$-space that winds in the counterclockwise direction around the BIC.
\begin{figure*}
	\centering
	\includegraphics[width=\linewidth]{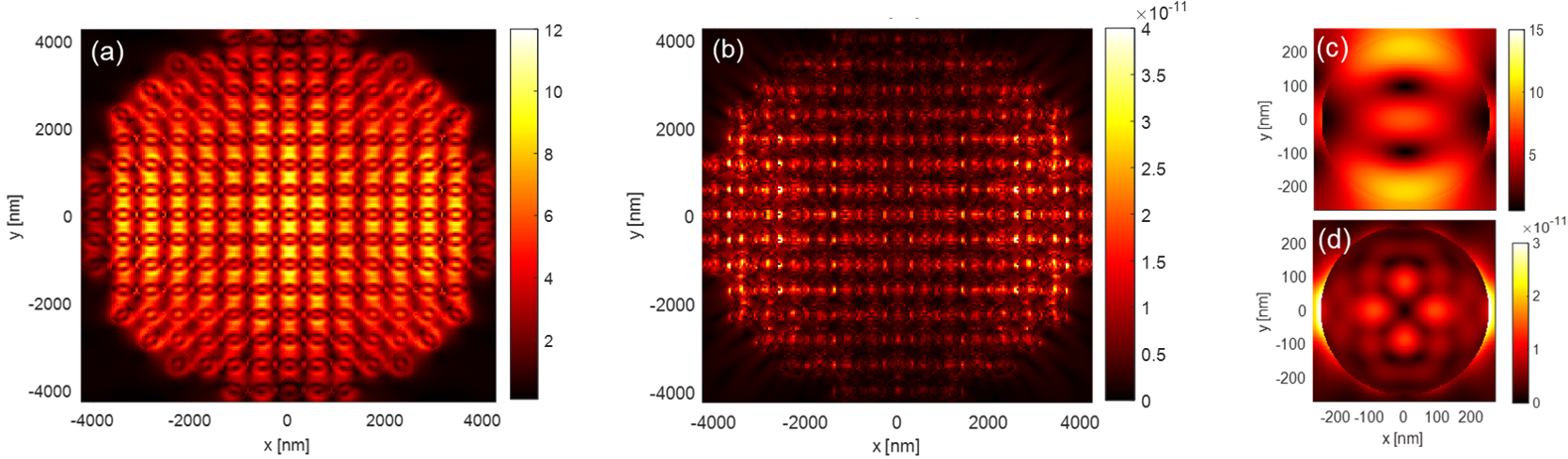}%[width=\columnwidth]
	\caption{(a), (b) Spatial distribution of the optical near-field in the cluster corresponding to the FF and SH, respectively. (c), (d) Spatial distribution of the optical near-field in the unit cell containing the central nanosphere of the cluster and corresponding to the FF and SH, respectively. The wavelengths of the FF and SH are $\lambda=\SI{1276}{\nm}$ and $\lambda=\SI{638}{\nm}$, respectively. In all panels, the optical near-field is computed in the plane defined by $z=0$.}
	\label{fig:fields_cl}
\end{figure*}

The results of these calculations are summarized in Fig.~\ref{fig:BSMP}. Thus, Figs.~\ref{fig:BSMP}(a) and \ref{fig:BSMP}(c) show the dependence of the resonance wavelength of the optical modes that exist at the FF and SH, respectively, on the period of the square lattice, $p$. The resonance wavelength of the modes is calculated at the $\Gamma$-point and the BIC bands are labelled by numbers ranging from \emph{1} to \emph{5}. At the FF there are three BIC bands with the following properties: the bands \textit{1} and \textit{3} are symmetry-protected BICs with diverging $Q$-factor at the $\Gamma$-point and topological charge $q=-1$ and $q=1$, respectively, whereas band \textit{2} is a double-degenerate BIC with topological charge $q=-1$, which is formed \textit{via} parameter tuning (its $Q$-factor diverges at the $\Gamma$-point when $p=\SI{602}{\nm}$). Moreover, at the SH there are two symmetry-protected BIC bands with diverging $Q$-factor at the $\Gamma$-point, labeled \textit{4} and \textit{5}, their topological charge being $q=1$ and $q=-1$, respectively. The spatial field profiles of the BIC bands at the FF and SH, calculated for $p=\SI{570}{\nm}$, are presented in Figs.~\ref{fig:BSMP}(b) and \ref{fig:BSMP}(d), respectively.

The dispersion of the BIC bands shown in Figs.~\ref{fig:BSMP}(a) and \ref{fig:BSMP}(c) suggests that it is possible to design a metasurface made of silicon nanospheres that possesses BIC modes both at the FF and FF. More specifically, it can be seen in these figures that the resonance wavelength of BICs \emph{1} and \emph{2} decreases with $p$, whereas the resonance wavelength of BICs \emph{4} and \emph{5} increases when $p$ increases. Therefore, there is a value of $p$ for which these BIC bands cross, period for which one expects a strong enhancement of the SHG. Whereas these conclusions hold for an infinite periodic metasurface, in the case of an optical cluster the finite-size and edge effects can change the linear and nonlinear optical response of the cluster, so that this physical configuration has to be further investigated.

To this end, we considered a cluster made of 177 silicon nanospheres placed in the $(x,y)$-plane at the vertices of a square lattice with period, $p$. The cluster is illuminated by an $x$-polarized plane wave propagating along the $z$-axis. For this configuration, we determined the scattering cross-sections at the FF and SH when the wavelength of the incident wave, $\lambda_{inc}$, varied from \SIrange{1270}{1300}{\nm} and the distance between adjacent nanospheres $p$ changed from \SIrange{570}{600}{\nm}. In these computations, we used 336 vector spherical wave functions per nanosphere, namely we set $l_{max}=12$.

We present the results of these calculations in Fig.~\ref{fig:scs_cl}, whereby we plot the dependence of the scattering cross-sections at the FF and SH on the wavelength of the incident wave and the distance between adjacent nanospheres. The plot in Fig.~\ref{fig:scs_cl}(a) clearly shows the resonant enhancement of the scattering cross-section at the FF upon excitation of the BICs \emph{1} and \emph{2}. Thus, as in the case of an infinite metasurface, we can see that the resonance wavelength of the two BIC bands decreases when the $p$ increases, with the BIC band \emph{2} being much less dispersive. Moreover, as expected, as the distance between adjacent nanospheres increases towards the value $p=\SI{602}{\nm}$, where the $Q$-factor diverges, the width of the BIC band \emph{2} decreases.

The existence of BICs at the FF and SH has a strong influence on the SHG, too, a fact clearly underlined by the dispersion map shown in Fig.~\ref{fig:scs_cl}(b). Thus, it can be inferred from this figure that the resonant excitation of BICs forming bands \emph{1} and \emph{2} results in a significant enhancement of the SHG from the optical cluster. This phenomenon is explained by the fact that the excitation of BICs at the FF leads to the enhancement of the local linear field, which in turn induces a marked enhancement of the nonlinear polarization at the SH. In addition, the existence of a BIC at the SH facilitates the radiation in the continuum of the SH, thus increasing the amount of the generated SH. Somewhat counterintuitive, it can be observed in Fig.~\ref{fig:scs_cl}(b) that the intensity of SHG corresponding to BIC \emph{2} is much larger than the one corresponding to BIC \emph{1}, despite the fact that the $Q$-factor of the former is merely $Q\simeq2,000$ whereas the $Q$-factor of the latter is $Q\simeq\num{1.1e9}$. This apparent contradiction is explained by the fact that the very large $Q$-factor of the BIC \emph{1} implies that only a small amount of light at the FF can couple into this resonance and therefore reduced amount of light is emitted at the SH.

Further insights into this double-resonance mechanism for the enhancement of the SHG in our optical cluster can be gleaned from the spatial distribution of the linear and nonlinear optical near-fields. To this end, we have determined the spatial distribution of the amplitude of the electric field at the FF and SH, the numerical results being plotted in Fig.~\ref{fig:fields_cl}. The spatial distribution of the linear field in the cluster, depicted in Fig.~\ref{fig:fields_cl}(a), indicates a strong interaction between the incoming plane wave and all nanospheres in the cluster, with an enhancement of the optical field by more than an order of magnitude being observed. At the SH, on the other hand, the nonlinear field is generated primarily around the nanospheres located near the edge of the cluster, as per Fig.~\ref{fig:fields_cl}(b). This interesting result can be understood on the basis of the symmetry properties of the cluster. In particular, the arrangement of the nanospheres form a structure that possesses inversion symmetry and therefore ``bulk'' SHG is suppressed. In other words, similarly to the suppression of bulk SHG in optical media whose crystalline atomic structure possesses inversion symmetry, so that SH is only generated at the surface, the SHG in the ``bulk'' of our planar cluster of silicon ``meta-atoms'' is suppressed being chiefly generated at the edge. Pushing this analogy even further would allow one to introduce an edge SHG, similarly to the surface SHG in nanoparticles made of centrosymmetric materials.

Finally, the field profiles presented in Fig.~\ref{fig:fields_cl} confirm that the linear field at the FF is primarily the result of the excitation of the BIC \emph{2}, whereas the nonlinear field at the SH is emitted chiefly from the BIC \emph{5}. To demonstrate this, we plot in Figs.~\ref{fig:fields_cl}(c) and \ref{fig:fields_cl}(d) the spatial distribution of the optical near-field in the unit cell containing the central nanosphere of the cluster and corresponding to the FF and SH, respectively. It can be easily seen that these spatial field profiles are very similar to the field distributions presented in Figs.~\ref{fig:BSMP}(b) and \ref{fig:BSMP}(d), corresponding to the BIC \emph{2} (the degenerate band) and BIC \emph{5} of a periodic, infinite metasurface. The small differences between these field profiles are possibly due to the finite-size and edge effects.

\section{Conclusion}\label{CONCL}
To summarize, in this manuscript we have extended the $\mathcal{T}$-matrix numerical formalism to the analysis of second-harmonic optical wave scattering from clusters of arbitrarily shaped particles made of centrosymmetric optical materials. The mathematical apparatus developed in this study is general in purpose, as it can be applied to an arbitrary distribution of dielectric, metallic, polaritonic, or semiconductor particles, as well as a mixture of such particles. In this work, we considered primarily particles with axial symmetry, a case in which the transfer matrix method performs the best due to the underlying spherical symmetry of the expansion functions employed. Despite this, our computational method performs well when applied to particle configurations that depart significantly from a spherical shape, although larger computational resources are required in such a case.

The numerical technique introduced here is derived from the integral representation of the optical fields at the fundamental frequency and second-harmonic, in conjunction with the null-field conditions within the extended boundary condition method. The corresponding integral operators are discretized using a vector spherical harmonic basis functions, thus establishing a matrix equation that relates the expansion coefficients of the unknown scattered fields to those of the known source fields. The nonlinear optical response of the optical cluster of particles is introduced in our analysis \textit{via} second-harmonic polarization sources located in the bulk and on the surface of the particles under investigation.

To illustrate the versatility and strength of our numerical method, we applied it to a series of relevant scattering configurations, including a single and multiple nanospheres, prolate spheroids, a dimer of spheroids, and a cluster of three particles. In all these cases the scatterers were made of either gold or silicon. In addition, we also considered a much more complex configuration, namely a large cluster of silicon nanospheres, and investigated the nonlinear optical interaction of bound-states in the continuum supported by the optical cluster.

These investigations revealed several advantages of our method as compared to the current state-of-the-art. In particular, our method requires only modest computational and memory resources as the optical scattered fields and scattering cross-sections are determined by a relatively small number of Fourier expansion coefficients. Moreover, since our method is not limited by the size of a computational domain, it can be used to solve the linear and nonlinear wave scattering problem in physical situations in which optical clusters of thousands or even tens of thousands of particles are considered. Given all these advantages and key valuable features that our method possesses, we expect it to prove to be instrumental in the design of optical clusters of nanoparticles, building blocks of linear and nonlinear metamaterials and metasurfaces with advanced functionalities, as well as other photonic nanodevices.

\section*{Acknowledgement}
This work was funded under the embedded CSE programme of the ARCHER2 UK National Supercomputing Service and used the ARCHER2 UK National Supercomputing Service  (http://www.archer2.ac.uk). We acknowledge the use of the UCL High Performance Computing Facility and associated support services in the completion of this work.

\appendix\section{Formulae for $\mathbf{Q}$-matrix elements}\label{MatElements}
\numberwithin{equation}{section}
In this Appendix we provide the mathematical expressions for the elements of the matrix $\mathbf{Q}^{\rho(1,3)}$, $\rho \in \{\omega, \Omega\}$, at FF and SH frequency, as well as the components of the excitation vector $\vect{g}^{\Omega(3)}$ at the SH. The corresponding elements of the matrix $\mathbf{Q}^{\rho(1,1)}$ and components of the vector $\vect{g}^{\Omega(1)}$ are obtained by replacing in these expressions the outgoing vector spherical harmonics, $(3)$, with the regular ones, $(1)$. Furthermore, we assume that a boundary mesh of the particle at hand is available, and defined as the union of $N_t$ non-overlapping flat triangular facets $\tilde{S}=\bigcup\limits_{p=1}^{N_t} S_p$.

The matrix $\mathbf{Q}^{\rho(1,3)}$ can be written as a combination of four submatrices, $\vect{Q}^{\rho,11}$, $\vect{Q}^{\rho,12}$, $\vect{Q}^{\rho,21}$, and $\vect{Q}^{\rho,22}$, \textit{i.e.}:
\begin{equation}\label{eq:Qmatrix_submat}
	\mathbf{Q}^{\rho(1,3)} =
	\begin{pmatrix}
		\vect{Q}^{\rho,11} & \vect{Q}^{\rho,12} \\ \vect{Q}^{\rho,21} & \vect{Q}^{\rho,22}
	\end{pmatrix},
\end{equation}
with the elements given by:
\begin{subequations}\label{eq:Qmatrix_elements}
	\begin{align}
Q^{\rho,11}_{\mu\nu}&= i(k_{e}^\rho)^2  \oiint_{\tilde{S}} \hat{\vect{n}} \cdot \left[\vect{M}_{\nu}^{(1)}(k_{i}^{\rho}\vect{r}) \times \vect{N}_{\bar{\mu}}^{(3)}(k_{e}^{\rho}\vect{r})\right]dS \nonumber\\+& i k_{e}^\rho k_{i}^\rho  \oiint_{\tilde{S}} \hat{\vect{n}} \cdot \left[\vect{N}_{\nu}^{(1)}(k_{i}^{\rho}\vect{r}) \times \vect{M}_{\bar{\mu}}^{(3)}(k_{e}^{\rho}\vect{r})\right] dS,\label{eq:Q_11}\\
Q^{\rho,12}_{\mu\nu}&=  i(k_{e}^\rho)^2  \oiint_{\tilde{S}} \hat{\vect{n}} \cdot \left[\vect{N}_{\nu}^{(1)}(k_{i}^{\rho}\vect{r}) \times \vect{N}_{\bar{\mu}}^{(3)}(k_{e}^{\rho}\vect{r})\right]dS \nonumber\\ +& i k_{e}^\rho k_{i}^\rho  \oiint_{\tilde{S}} \hat{\vect{n}} \cdot \left[\vect{M}_{\nu}^{(1)}(k_{i}^{\rho}\vect{r}) \times \vect{M}_{\bar{\mu}}^{(3)}(k_{e}^{\rho}\vect{r})\right] dS,\label{eq:Q_12}\\
Q^{\rho,21}_{\mu\nu}&=  i(k_{e}^\rho)^2  \oiint_{\tilde{S}} \hat{\vect{n}} \cdot \left[\vect{M}_{\nu}^{(1)}(k_{i}^{\rho}\vect{r}) \times \vect{M}_{\bar{\mu}}^{(3)}(k_{e}^{\rho}\vect{r})\right]dS \nonumber\\ +& i k_{e}^\rho k_{i}^\rho  \oiint_{\tilde{S}} \hat{\vect{n}} \cdot \left[\vect{N}_{\nu}^{(1)}(k_{i}^{\rho}\vect{r}) \times \vect{N}_{\bar{\mu}}^{(3)}(k_{e}^{\rho}\vect{r})\right] dS,\label{eq:Q_21}\\
Q^{\rho,22}_{\mu\nu}&=  i(k_{e}^\rho)^2  \oiint_{\tilde{S}} \hat{\vect{n}} \cdot \left[\vect{N}_{\nu}^{(1)}(k_{i}^{\rho}\vect{r}) \times \vect{M}_{\bar{\mu}}^{(3)}(k_{e}^{\rho}\vect{r})\right]dS \nonumber\\ +& i k_{e}^\rho k_{i}^\rho  \oiint_{\tilde{S}} \hat{\vect{n}} \cdot \left[\vect{M}_{\nu}^{(1)}(k_{i}^{\rho}\vect{r}) \times \vect{N}_{\bar{\mu}}^{(3)}(k_{e}^{\rho}\vect{r})\right] dS,\label{eq:Q_22}
	\end{align}
\end{subequations}
where $\nu$ and $\bar{\mu}$ are the combined indices $(l, m)$ and $(l^{\prime}, -m^{\prime})$, respectively.

The components of the excitation vector $\vect{g}^{\Omega(3)}$ are defined by surface integrals of products between tangential source currents and VSWFs. More specifically, the vector $\vect{g}^{\Omega(3)}$ can be decomposed into a stack of vectors $\vect{g}_{\mu}^{\Omega}$ defined as:
\begin{equation}\label{eq:gvect_subvect}
\vect{g}_{\mu}^{\Omega}=\oiint_{\tilde{S}}\bm{\mathcal{C}}^{\Omega}(\vect{r})\cdot
\begin{pmatrix}
\vect{M}_{\bar{\mu}}^{(3)}(k_{e}^{\Omega}\vect{r})\\
\vect{N}_{\bar{\mu}}^{(3)}(k_{e}^{\Omega}\vect{r})
\end{pmatrix}
dS,
\end{equation}
where the $2\times2$ matrix $\bm{\mathcal{C}}^{\Omega}(\vect{r})$ is defined in terms of the electric and magnetic currents, $\bm{\mathcal{J}}^{\Omega}(\vect{r})$ and $\bm{\mathcal{M}}^{\Omega}(\vect{r})$, respectively, and the internal field $\vect{E}_{i,p}^{\Omega}(\vect{r})$:
\begin{subequations}\label{eq:gvector_elements}
\begin{align}
&\bm{\mathcal{C}}^{\Omega}_{11}=\bm{\mathcal{C}}^{\Omega}_{22}=\mu_0\Omega k_{e}^\Omega\bm{\mathcal{J}}^{\Omega},\\
&\bm{\mathcal{C}}^{\Omega}_{12}=\bm{\mathcal{C}}^{\Omega}_{21}=i(k_{e}^\Omega)^2 \left(\bm{\mathcal{M}}^{\Omega} - \hat{\mathbf{n}} \times \vect{E}_{i,p}^{\Omega}\right).
\end{align}
\end{subequations}

The integrals in Eq.~\eqref{eq:gvect_subvect} are of three different types, namely
\begin{subequations}\label{eq:integral_parts_g}
	\begin{align}
		&I_1 = \mu_0\Omega k_{e}^\Omega\oiint_{\tilde{S}} \bm{\mathcal{J}}^{\Omega}(\vect{r}) \cdot \vect{M}_{\bar{\mu}}^{(3)}(k_{e}^{\Omega}\vect{r})\,dS, \label{eq:I_1}\\
		&I_1^{\prime} = \mu_0\Omega k_{e}^\Omega\oiint_{\tilde{S}} \bm{\mathcal{J}}^{\Omega}(\vect{r}) \cdot \vect{N}_{\bar{\mu}}^{(3)}(k_{e}^{\Omega}\vect{r})\,dS, \label{eq:I_1prime}\\
		&I_2 =i(k_{e}^\Omega)^2 \oiint_{\tilde{S}} \bm{\mathcal{M}}^{\Omega}(\vect{r}) \cdot \vect{N}_{\bar{\mu}}^{(3)}(k_{e}^{\Omega}\vect{r})\,dS, \label{eq:I_2}\\
		&I_2^{\prime} =i(k_{e}^\Omega)^2 \oiint_{\tilde{S}} \bm{\mathcal{M}}^{\Omega}(\vect{r}) \cdot \vect{M}_{\bar{\mu}}^{(3)}(k_{e}^{\Omega}\vect{r})\,dS, \label{eq:I_2prime}\\
		&I_3 =i(k_{e}^\Omega)^2\oiint_{\tilde{S}} \left[\hat{\mathbf{n}} \times \vect{E}_{i,p}^{\Omega}(\mathbf{r})\right] \cdot \vect{N}_{\bar{\mu}}^{(3)}(k_{e}^{\Omega}\vect{r})\,dS, \label{eq:I_3}\\
		&I_3^{\prime} =i(k_{e}^\Omega)^2\oiint_{\tilde{S}} \left[\hat{\mathbf{n}} \times \vect{E}_{i,p}^{\Omega}(\mathbf{r})\right] \cdot \vect{M}_{\bar{\mu}}^{(3)}(k_{e}^{\Omega}\vect{r})\,dS. \label{eq:I_3prime}
	\end{align}
\end{subequations}

First, let us consider the evaluation of the integral $I_1$. With the help of Eqs.~\eqref{eq:NonlCurr_a} and \eqref{eq:SHSurfPolarization_Comp}, we can write:
\begin{align}\label{eq:I_1_eval}
I_1 =& 2 i \chi_{s,\parallel\perp\parallel}^{(2)}k_{e}^\Omega(k_{0}^\Omega)^2 \oiint_{\tilde{S}} E_{n}^{\omega}(\vect{r})\\
& \cdot \left\{\left[\hat{\mathbf{n}} \times \hat{\mathbf{n}} \times \vect{E}_{t}^{\omega} (\vect{r})\right] \cdot \vect{M}_{\bar{\mu}}^{(3)}(k_{e}^{\Omega}\vect{r})\right\}dS\nonumber,
\end{align}
where the fields are evaluated on the inner side of the surface of the particle and $k_{0}^\Omega$ is the wave number in vacuum at frequency $\Omega$. Written in this form, the integral $I_1$ is suitable for the numerical evaluation.

Now we turn our attention to the integral $I_2$ and use Eq.~\eqref{eq:NonlCurr_b} defining the nonlinear magnetic surface current in terms of the normal component of the surface polarization density source, $P_{n,s}^{\Omega}(\vect{r})=\hat{\vect{n}} \cdot \bm{\vect{P}}_{s}^{\Omega}(\vect{r})$:
\begin{equation}\label{eq:I_2_eval}
	I_2 = \frac{i(k_{e}^\Omega)^2}{\epsilon_0}\oiint_{\tilde{S}} \left[\hat{\mathbf{n}} \times \nabla_S P_{s,n}^{\Omega}(\vect{r})\right] \cdot \vect{N}_{\bar{\mu}}^{(3)}(k_{e}^{\Omega}\vect{r})\,dS.
\end{equation}
We break this integral over the triangular patches as:
\begin{align}\label{eq:I_2_eval_trian}
&I_2 = \frac{i(k_{e}^\Omega)^2}{\epsilon_0} \sum\limits_{p=1}^{N_t}\left\{\oint_{C_p} P_{s,n}^{\Omega}(\vect{r}) \vect{N}_{\bar{\mu}}^{(3)}(k_{e}^{\Omega}\vect{r}) \cdot d\mathbf{l}\right.\nonumber\\& \left.- \iint_{S_p}  P_{s,n}^{\Omega}(\vect{r}) \left[\nabla_S \times \vect{N}_{\bar{\mu}}^{(3)}(k_{e}^{\Omega}\vect{r})\right] \cdot \hat{\mathbf{n}}\,dS\right\}.
\end{align}
In this equation, $C_p=\partial S_p$ is the boundary of the $p$th triangle in the mesh and the direction of integration along the path $C_p$ is chosen in such a way that the right-hand-screw rule gives the direction of the outwardly oriented normal, $\hat{\mathbf{n}}$. To derive Eq.~\eqref{eq:I_2_eval_trian} from Eq.~\eqref{eq:I_2_eval} one first uses the vector calculus relation $\left[\hat{\mathbf{n}} \times \nabla_S \psi(\vect{r})\right] \cdot \vect{A}(\vect{r})=\left\{\nabla_S\times[\psi(\vect{r})\vect{A}(\vect{r})] -\psi(\vect{r})\nabla_S\times\vect{A}(\vect{r})\right\}\cdot\hat{\mathbf{n}}$, where $\psi(\vect{r})$ and $\vect{A}(\vect{r})$ are arbitrary smooth scalar and vector functions, respectively, and then employs the Stokes theorem.

The numerical evaluation of the integral $I_2$ is greatly simplified by the realization that the contour integral on the right hand side of Eq.~\eqref{eq:I_2_eval_trian} is equal to zero when it is evaluated on a closed mesh without cracks and slits. Under these circumstances, the final expression for the integral $I_2$ is:
\begin{equation}\label{eq:I_2_eval_final}
	I_2 = -i(k_{e}^\Omega)^3\chi_{s,\perp\perp\perp}^{(2)}  \oiint_{\tilde{S}} [E_{n}^{\omega}(\vect{r})]^2 \vect{M}_{\bar{\mu}}^{(3)}(k_{e}^{\Omega}\vect{r}) \cdot \hat{\mathbf{n}}\,dS.
\end{equation}
In the derivation of this expression we took advantage of the curl identities in Eqs.~\eqref{eq:regularVSWF} satisfied by the VSWFs, as well as the fact that the surface tensor component $\chi_{s,\perp\parallel\parallel}^{(2)}$ is vanishingly small and thus its contribution to the SHG can be disregarded. As such, the normal component of the surface polarization density is: $P_{s,n}^{\Omega}(\vect{r}) = \epsilon_0 \chi_{s,\perp\perp\perp}^{(2)} [E_{n}^{\omega}(\vect{r}_s)]^2$.

The last integral, $I_3$, is written as:
\begin{equation}\label{eq:I3_eval}
I_3 =\frac{i(k_{e}^\Omega)^2 \epsilon_0 \gamma}{\epsilon_i^{\Omega}}\oiint_{\tilde{S}} \left\{\hat{\mathbf{n}} \times \nabla \left[\vect{E}_{i}^{\omega}(\mathbf{r}) \cdot
\vect{E}_{i}^{\omega}(\mathbf{r})\right]\right\} \cdot \vect{N}_{\bar{\mu}}^{(3)}(k_{e}^{\Omega}\vect{r})\,dS,
\end{equation}
where we have used the particular electric field solution given by Eq.~\eqref{eq:PartSH_E} of the inhomogeneous system of Maxwell equations at the SH. This integral can be calculated using a similar method to that used to calculate $I_{2}$, the end result being:
\begin{equation}\label{eq:I3_finale}
I_3 =-\frac{i(k_{e}^\Omega)^3 \epsilon_0 \gamma}{\epsilon_i^{\Omega}}\oiint_{\tilde{S}} \left[\vect{E}_{i}^{\omega}(\mathbf{r}) \cdot \vect{E}_{i}^{\omega}(\mathbf{r})\right] \left[\vect{M}_{\bar{\mu}}^{(3)}(k_{e}^{\Omega}\vect{r}) \cdot \hat{\mathbf{n}}\right]\,dS.
\end{equation}

Finally, the primed integrals, $I_1^{\prime}$, $I_2^{\prime}$, and $I_3^{\prime}$, can be obtained from $I_1$, $I_2$, and $I_3$, respectively, by interchanging the spherical harmonics $\vect{M}_{\bar{\mu}}^{(3)}(k_{e}^{\Omega}\vect{r})\leftrightarrow\vect{N}_{\bar{\mu}}^{(3)}(k_{e}^{\Omega}\vect{r})$ in Eqs.~\eqref{eq:I_1_eval}, \eqref{eq:I_2_eval_final}, and \eqref{eq:I3_finale}. All of the integrals can be computed numerically using Gauss-Legendre quadrature in six points.

\section{Definition and formulae for vector spherical harmonics}\label{Maths}
\numberwithin{equation}{section}
In this Appendix we introduce the definition and main properties of vector spherical harmonics
(VSHs) used in our study to represent the electromagnetic fields, as well as to expand the plane waves.
\subsection*{Definition of vector spherical wave functions}
Given the unit vector $\hat{\vect{r}} = \vect{r}/r$, where $\vect{r}$ is the position vector of a
point, $P$, we define three sets of mutually orthogonal VSHs: longitudinal VSHs, $\vect{Y}_{lm}$,
which are oriented along $\hat{\vect{r}}$, and two sets of transverse VSHs, $\vect{X}_{lm}$ and
$\vect{Z}_{lm}$, oriented perpendicular to $\hat{\vect{r}}$ \cite{Varshalovich,Stout2}. The triplet
$\{\vect{X}_{lm},\vect{Y}_{lm},\vect{Z}_{lm}\}$ of VSHs form a right-handed system of vectors and
are defined as:
\begin{subequations}
	\begin{align}
		&\vect{Y}_{lm}(\theta,\phi) =  Y_{lm}(\theta,\phi)\hat{\vect{r}},\label{eq:Xm1}\\
		&\vect{X}_{lm}(\theta,\phi) = \frac{1}{\sqrt{l(l+1)}} \hat{\vect{r}} \times \nabla_{\hat{\vect{r}}}
		Y_{lm}(\theta,\phi),\label{eq:Xzero}\\
		&\vect{Z}_{lm}(\theta,\phi) = \frac{1}{\sqrt{l(l+1)}} \nabla_{\hat{\vect{r}}}
		Y_{lm}(\theta,\phi).\label{eq:Xp1}
	\end{align}
	\label{eq:X}
\end{subequations}

Furthermore, we define the regular VSWFs $\vect{M}_{lm}^{(1)}$ and $\vect{N}_{lm}^{(1)}$ as
follows:
\begin{subequations}\label{eq:regularVSWF}
	\begin{align}
		&\vect{M}_{lm}^{(1)}(kr,\theta,\phi) = -j_l(kr) \vect{X}_{lm}, \\
		&\vect{N}_{lm}^{(1)}(kr,\theta,\phi) = \frac{1}{k} \nabla \times \vect{M}_{lm}^{(1)}, \\
		&\vect{M}_{lm}^{(1)}(kr,\theta,\phi) = \frac{1}{k} \nabla \times \vect{N}_{lm}^{(1)},
	\end{align}
\end{subequations}
where $k$ is the wave number calculated at the particular frequency in the medium of choice and
$j_l(kr)$ are spherical Bessel functions of the first kind and order $l$. The radiative (outgoing)
VSWFs $\vect{M}_{lm}^{(3)}$ and $\vect{N}_{lm}^{(3)}$ are defined by substituting the spherical
Bessel functions in \eqref{eq:regularVSWF} with (outgoing) spherical Hankel functions,
$h_l^{(1)}(kr)$, of the first kind and order $l$.

Scalar spherical harmonics can be defined in terms of associated Legendre functions $P_l^m$ as
\cite{Stout2}:
\begin{equation}
	Y_{lm}(\theta, \phi) = \gamma_{lm} \sqrt{l(l+1)} P_l^m(\cos \theta) e^{im\phi},
	\label{eq:ScaSphHar}
\end{equation}
where $\gamma_{lm}$ is the normalization coefficient given by:
\begin{equation}
	\gamma_{lm} = \sqrt{\frac{(2l + 1)(l - m)!}{4 \pi l(l + 1)(l + m)!}}. \label{eq:Normcoeff}
\end{equation}

In these definitions the scalar spherical harmonic functions $Y_{lm}(\hat{\mathbf{r}})$ and the
three kinds of VSHs are normalized to unity over a sphere:
\begin{subequations}\label{eq:normSphHarm}
	\begin{align}		&\iint_{\Omega_{\hat{\mathbf{r}}}}Y_{lm}^{*}(\hat{\mathbf{r}})Y_{l^{\prime}m^{\prime}}(\hat{\mathbf{r}})d\Omega_{\hat{\mathbf{r}}}
		= \delta_{ll^{\prime}}\delta_{mm^{\prime}},\\
		&\iint_{\Omega_{\hat{\mathbf{r}}}}\mathbf{A}_{lm}^{*}(\hat{\mathbf{r}})\cdot
		\mathbf{B}_{l^{\prime}m^{\prime}}(\hat{\mathbf{r}})d\Omega_{\hat{\mathbf{r}}} =
		\delta_{ll^{\prime}}\delta_{mm^{\prime}}\delta_{AB},
	\end{align}
\end{subequations}
where $\mathbf{A}$ and $\mathbf{B}$ can be any of the functions $\mathbf{X}$, $\mathbf{Y}$, and
$\mathbf{Z}$.

Finally, the normalized VSHs can be defined in terms of associated Legendre functions as:
\begin{subequations}\label{eq:normLegd}
	\begin{align}
\vect{Y}_{lm}(\theta,\phi)=\,& \gamma_{lm}\sqrt{l(l+1)} P_l^m (\cos \theta)
		e^{im\phi}\hat{\vect{r}}, \label{eq:Xm1Legd}\\
\vect{X}_{lm}(\theta,\phi)=\,& \gamma_{lm} \left[-\frac{im}{\sin \theta} P_l^m (\cos
		\theta)e^{im\phi} \hat{\bm{\theta}}\right.\nonumber\\ &+\left.\frac{d}{d \theta} P_l^m (\cos \theta)e^{im\phi}
		\hat{\bm{\phi}} \right], \label{eq:XzeroLegd}\\
\vect{Z}_{lm}(\theta, \phi)=\,& \gamma_{lm} \left[ \frac{d}{d \theta} P_l^m (\cos \theta)e^{im\phi}
		\hat{\bm{\theta}}\right.\nonumber\\ &+\left.\frac{im}{\sin \theta} P_l^m (\cos \theta)e^{im\phi}
		\hat{\bm{\phi}} \right]. \label{eq:Xp1Legd}
	\end{align}
\end{subequations}

\subsection*{Expansion of plane waves in series of vector spherical wave functions}
The series expansion of a plane wave in terms of VSWFs is given in \eqref{eq:IncidentVSWFE} and \eqref{eq:IncidentVSWFH}, with
the expansion coefficients $q_{lm}^{\omega}$ and $p_{lm}^{\omega}$ calculated as follows
\cite{Stout2}:
\begin{subequations}\label{eq:PLWexpCoeff}
	\begin{align}
		q_{lm}^{\omega} &= -4\pi i^l \vect{X}_{lm}^{*}(\hat{\vect{k}}_{inc}) \cdot \hat{\vect{e}}_{inc},\\
		p_{lm}^{\omega} &= -4\pi i^{l+1} \vect{Z}_{lm}^{*}(\hat{\vect{k}}_{inc}) \cdot
		\hat{\vect{e}}_{inc},
	\end{align}
\end{subequations}
where the symbol ``$*$'' denotes complex conjugation, $\hat{\vect{k}}_{inc}$ is the unit vector in
the direction of the wave propagation, and $\hat{\vect{e}}_{inc}$ is the polarization direction of
the plane wave.

%Manual citation list

%\bibliography{Notes}

\begin{thebibliography}{1}
    % Intoroduction
    \bibitem[Boyd(1992)]{Boyd} R. W. Boyd, \textit{Nonlinear Optics} (Academic Press, 1992).
    %
    \bibitem[Ray(2003)]{Ray} P. C. Ray, Size and Shape Dependent Second Order Nonlinear Optical Properties of Nanomaterials and Their Application in Biological and Chemical Sensing, Chem. Rev. \textbf{110}, 5332 (2010).
    %
    \bibitem[Hatab(2010)]{Hatab} N. A. Hatab, C. H. Hsueh, A. L. Gaddis, S. T. Retterer, J. H. Li, G. Eres, Z. Zhang, and B. Gu, Free-Standing Optical Gold Bowtie Nanoantenna with Variable Gap Size for Enhanced Raman Spectroscopy, Nano Lett. \textbf{10}, 4952 (2010).
    %
    \bibitem[Butet(2015)]{Butet} J. Butet, P. F. Brevet, and O. J. F. Martin, Optical second harmonic generation in plasmonic nanostructures: from fundamental principles to advanced applications. ACS Nano \textbf{9}, 10545 (2015).
    %
    \bibitem[Panoiu(2018)]{Panoiu} N. C. Panoiu, W. E. I. Sha, D. Y. Lei, and G. C. Li, Nonlinear optics in plasmonic nanostructures, J. Opt. \textbf{20}, 083001 (2018).
    %
    \bibitem[Homola(2008)]{Homola} J. Homola, Surface plasmon resonance sensors for detection of chemical and biological species, Chem. Rev. \textbf{108}, 462 (2008).
    %
    \bibitem[Pendry(1999)]{Pendry} J. B. Pendry, A. J. Holden, D. J. Robbins, and W. J. Stewart, Magnetism from conductors and enhanced nonlinear phenomena, IEEE Trans. Microw. Theory Tech. \textbf{47}, 2075 (1999).
    %
    \bibitem[Liu(2008)]{Liu} N. Liu, H. Guo, L. Fu, S. Kaiser, H. Schweizer, and H. Giessen, Three-dimensional photonic metamaterials at optical frequencies, Nat. Mater. \textbf{7}, 31 (2008).
    %
    \bibitem[Ziolkowski(2004)]{Ziolkowski} R. W. Ziolkowski, Propagation in and scattering from a matched metamaterial having a zero index of refraction, Phys. Rev. E \textbf{70}, 046608 (2004).
    %
    \bibitem[Canfield(2006)]{Canfield} B. K. Canfield, S. Kujala, K. Laiho, K. Jefimovs, J. Turunen, and M. Kauranen, Chirality arising from small defects in gold nanoparticle arrays, Opt. Express \textbf{14}, 950 (2006).
    %
    \bibitem[Valev(2011)]{Valev} V. K. Valev, X. Zheng, C. G. Biris, A. V. Silhanek, V. Volskiy, B. De Clercq, O. A. Aktsipetrov, M. Ameloot, N. C. Panoiu, G. A. E. Vandenbosch, and V. V. Moshchalkov, The origin of second harmonic generation hotspots in chiral optical metamaterials, Opt. Mater. Express \textbf{1}, 36 (2011).
    %
    \bibitem[Malkiel(2018)]{Malkiel} I. Malkiel, M. Mrejen, A. Nagler, U. Arieli, L. Wolf, and H. Suchowski, Plasmonic nanostructure design and characterization via Deep Learning, Light: Science \& Applications \textbf{7}, 60 (2018).
    %
    \bibitem[Molesky(2018)]{Molesky} S. Molesky, Z. Lin, A. Y. Piggott, W. Jin, J. Vuckovic, and A. W. Rodriguez, Inverse design in nanophotonics, Nat. Photon. \textbf{12}, 659 (2018).
    %
    \bibitem[Kudyshev(2020)]{Kudyshev} Z. A. Kudyshev, A. V. Kildishev, V. M. Shalaev, and A. Boltasseva, Machine-learning-assisted metasurface design for high efficiency thermal emitter optimization, Appl. Phys. Rev. \textbf{7}, 021407 (2020).
    %
    \bibitem[Shen(2003)]{Shen} Y. R. Shen, \textit{The Principles of Nonlinear Optics} (Wiley, 2003).
    %
    \bibitem[Biris(2010)]{Biris} C. G. Biris and N. C. Panoiu, Second harmonic generation in metamaterials based on homogeneous centrosymmetric nanowires, Phys. Rev. B \textbf{81}, 195102 (2010).
    %
    \bibitem[Pavlyukh(2004)]{Pavlyukh} Y. Pavlyukh and W. Hubner, Nonlinear Mie scattering from spherical particles, Phys. Rev. B \textbf{70}, 245434 (2004).
    %
    \bibitem[Beer1(2007)]{Beer1} A. G. F. de Beer and S. Roke, Nonlinear Mie theory for second-harmonic and sum-frequency scattering, Phys. Rev. B \textbf{79}, 155420 (2007).
    %
    \bibitem[Forestiere1(2014)]{Forestiere1} A. Capretti, C. Forestiere, L. Dal Negro, and G. Miano, Full-wave analytical solution of second-harmonic generation in metal nanospheres, Plasmonics \textbf{9}, 151 (2014).
    %
    \bibitem[Xu(2012)]{Xu} J. Xu and X. Zhang, Second harmonic generation in three-dimensional structures based on homogeneous centrosymmetric metallic spheres, Opt. Express \textbf{20}, 1668 (2012).
    %
    \bibitem[Sekulic(2021)]{Sekulic1} I. Sekulic, J. W. You, and N. C. Panoiu, \textit{T}-matrix method for calculation of second-harmonic generation in clusters of spherical particles, J. Quant. Spectrosc. Radiat. Transfer \textbf{268}, 107643 (2021).
    %
    \bibitem[Silvester(1996)]{Silvester} P. P. Silvester and R. L. Ferrari, \textit{Finite Elements for Electrical Engineers}, (Cambridge University Press, 1996).
    %
    \bibitem[Jin(2002)]{Jin} J. M. Jin, \textit{The Finite Element Method in Electromagnetics}, (John
    Wiley \& Sons, 2002).
    %
    \bibitem[Yee(1966)]{Yee} K. Yee, Numerical solution of initial boundary value problems involving Maxwell's equations in isotropic media, IEEE Trans. Antennas Propag. \textbf{AP-14}, 302 (1966).
    %
    \bibitem[Taflove(2005)]{Taflove} A. Taflove and S. C. Hagness, \textit{Computational Electrodynamics: The Finite-Difference Time-Domain Method}, (Artech House, 2005).
    %
    \bibitem[Mayergoyz(2005)]{Mayergoyz} I. D. Mayergoyz, D. R. Fredkin, and Z. Zhang, Electrostatic (plasmon) resonances in nanoparticles, Phys. Rev. B \textbf{72}, 155412 (2005).
    %
    \bibitem[Hohenester1(2005)]{Hohenester1} U. Hohenester and J. Krenn, Surface plasmon resonances of single and coupled metallic nanoparticles: A boundary integral method approach, Phys. Rev. B \textbf{72}, 195429 (2005).
    %
    \bibitem[Poggio(1973)]{Poggio} A. J. Poggio and E. K. Miller, Integral equation solutions of three-dimensional scattering problems, in \textit{Computer Techniques for Electromagnetics}, edited by R. Mittra (Pergamon Press, Oxford, 1973).
    %
    \bibitem[Wu(1977)]{Wu} T. K. Wu and L. L. Tsai, Scattering from arbitrarily-shaped lossy dielectric bodies of revolution, Radio Sci. \textbf{12}, 709 (1977).
    %
     \bibitem[Makitalo(2011)]{Makitalo} J. Makitalo, S. Suuriniemi, and M. Kauranen, Boundary element
    method for surface nonlinear optics of nanoparticles, Opt. Express \textbf{19}, 23386 (2011).
    %
    \bibitem[Forestiere2(2013)]{Forestiere2} C. Forestiere, A. Capretti, and G. Miano, Surface integral method for second harmonic generation in metal nanoparticles including both local-surface and nonlocal-bulk sources, J. Opt. Soc. Am. B \textbf{30}, 2355 (2013).
    %
    \bibitem[Harrington(1968)]{Harrington} R. F. Harrington, \textit{Field Computation by Moment Method}, (Macmillan 1968).
    %
    \bibitem[Rao(1982)]{Rao} S. M. Rao, D. R. Wilton, and A. W. Glisson, Electromagnetic scattering by surfaces of arbitrary shape, IEEE Trans. Antennas Propag. \textbf{30}, 409 (1982).
    %
    \bibitem[Song(1997)]{Song} J. Song, C. C. Lu, and W. C.  Chew, Multilevel fast multipole algorithm for electromagnetic scattering by large complex objects, IEEE Trans. Antennas Propag. \textbf{45}, 1488 (1997).
    %
    \bibitem[Bebendorf(2000)]{Bebendorf} M. Bebendorf, Approximation of boundary element matrices, Numer. Math. \textbf{86}, 565 (2000).
    %
    \bibitem[Varshalovich(1988)]{Varshalovich} D. A. Varshalovich, A. N. Moskalev, and V. K. Khersonskii, \textit{Quantum Theory of Angular Momentum}, (World Scientific Publishing 1988).
    %
    \bibitem[Jackson(1999)]{Jackson} J. D. Jackson, \textit{Classical Electrodynamics}, (John
    Wiley \& Sons, 1999).
    %
    \bibitem[Waterman1(1965)]{Waterman1} P. C. Waterman, Matrix formulation of electromagnetic scattering,
    Proc. IEEE \textbf{53}, 805 (1965).
    %
    \bibitem[Waterman2(1969)]{Waterman2} P. C. Waterman, Scattering by dielectric obstacles, Alta Freq. \textbf{38}, 348 (1969).
    %
    \bibitem[Barber(1975)]{Barber} P. Barber and C. Yeh, Scattering of electromagnetic waves by arbitrarily shaped dielectric bodies, Appl. Opt. \textbf{14}, 2864 (1975).
    %
    \bibitem[Mishchenko1(1996)]{Mishchenko0} M. I. Mishchenko, L. D. Travis, and D. W. Mackowski, \textit{T}-matrix computations of light scattering by nonspherical particles: a review, J. Quant. Spectrosc. Radiat. Transfer \textbf{55}, 535 (1996).
    %
    \bibitem[Mishchenko1(2007)]{Mishchenko1} M. I. Mishchenko, G. Videen, V. A. Babenko, N. G. Khlebtsov, and T. Wriedt, Comprehensive \textit{T}-matrix reference database: A 2004-06 update, J. Quant. Spectrosc. Radiat. Transfer \textbf{106}, 304 (2007).
    %
    \bibitem[Cruzan(1962)]{Cruzan} O. R. Cruzan, Translational addition theorems for spherical vector wave functions, Quart. Appl. Math. \textbf{20}, 33 (1962).
    %
    \bibitem[Mackowski(1996)]{Mackowski} D. W. Mackowski and M. I. Mishchenko, Calculation of the \textit{T}-matrix and the scattering matrix for ensembles of spheres, J. Opt. Soc. Am. A \textbf{13}, 2266 (1996).
    %
    \bibitem[Stout1(2002)]{Stout1} B. Stout, J. C. Auger, and J. Lafait, A transfer matrix approach to local field calculations in multiple-scattering problems, J. Mod. Opt. \textbf{49}, 2129 (2002).
    %
    \bibitem[Timbrell(2018)]{Timbrell} D. Timbrell, J. W. You, Y. S. Kivshar, and N. C Panoiu, A comparative analysis of surface and bulk contributions to second-harmonic generation in centrosymmetric nanoparticles, Sci. Rep. \textbf{8}, 3586 (2018).
    %
    \bibitem[Dadap(2004)]{Dadap} J. I. Dadap, J. Shan, and T. F. Heinz, Theory of optical second-harmonic generation from a sphere of centrosymmetric material: small-particle limit, J. Opt. Soc. Am. B \textbf{21}, 1328 (2004).
    %
    \bibitem[Bloembergen(1968)]{Bloembergen} N. Bloembergen, R. K. Chang, S. S. Jha, and C. H. Lee, Optical second-harmonic generation in reflection from media with inversion symmetry, Phys. Rev. \textbf{174}, 813 (1968).
    %
    \bibitem[Sipe1(1980)]{Sipe1} J. E. Sipe, V. C. Y. So, M. Fukui, and G. I. Stegeman, Analysis of second-harmonic generation at metal surfaces, Phys. Rev. B \textbf{21}, 4389 (1980).
    %
    \bibitem[Tsang(1985)]{Tsang} L. Tsang, J. A. Kong, and R. T. Shin, \textit{Theory of Microwave Remote Sensing}, (Wiley, 1985).
    %
    \bibitem[Stratton(1939)]{Stratton} J. A. Stratton and L. J. Chu, Diffraction theory of electromagnetic waves, Phys. Rev. \textbf{56}, 99 (1939).
    %
     \bibitem[Chew(2009)]{Chew} W. C. Chew, M. S. Tong, and B. Hu, \textit{Integral Equation Methods for Electromagnetic and Elastic Waves}, (Morgan \& Claypool, 2009).
     %
    \bibitem[Morse(1953)]{Morse} P. Morse and H. Feshbach, \textit{Methods of Theoretical Physics}, (McGraw-Hill, 1953).
    %
    \bibitem[Sekulic(2019)]{Sekulic2} I. Sekulic, D. C. Tzarouchis, P. Yla-Oijala, E. Ubeda, and J. M. Rius, Enhanced discretization of surface integral equations for resonant scattering analysis of sharp-edged plasmonic nanoparticles, Phys. Rev. B \textbf{99}, 165417 (2019).
    %
    \bibitem[Heinz(1991)]{Heinz} T. F. Heinz, in \textit{Nonlinear Surface Electromagnetic Phenomena}, edited by H. E. Ponath and G. I. Stegeman (Elsevier, Amsterdam, 1991), p. 353.
    %
   \bibitem[Stout2(2001)]{Stout2} B. Stout, J. C. Auger, and J. Lafait, Individual and aggregate scattering matrices and cross-sections: conservation laws and reciprocity, J. Mod. Opt. \textbf{48}, 2105 (2001).
   %
   \bibitem{optimet} OPTIMET-3D, github.com/OPTIMET/OPTIMET.
   %
   \bibitem[Mie1(1908)]{Mie1}G. Mie, Beitrage zur optik trüber medien, speziell kolloidaler metallösungen, Ann. Phys. (Berlin) \textbf{330}, 377 (1908).
   %
   \bibitem{CSTStudio}CST Studio\textsuperscript{\textregistered}, www.cst.com.
    %
    \bibitem[Schinke(2015)]{Schinke} C. Schinke, P. C. Peest, J. Schmidt, R. Brendel, K. Bothe, M. R. Vogt, I. Kroger, S. Winter, A. Schirmacher, S. Lim, H. T. Nguyen, and D. MacDonald, Uncertainty analysis for the coefficient of band-to-band absorption of crystalline silicon, AIP Advances \textbf{5}, 067168 (2015).
    %
    \bibitem[Johnson(1972)]{Johnson} P. B. Johnson and R. W. Christy, Optical constants of the noble metals, Phys. Rev. B \textbf{6}, 4370 (1972).
    %
    \bibitem[Rudnick(1971)]{Rudnick} J. Rudnick and E. A. Stern, Second-harmonic radiation from metal surfaces, Phys. Rev. B \textbf{4}, 4274 (1971).
    %
    \bibitem[Falasconi(2001)]{Falasconi} M. Falasconi, L. C. Andreani, A. M. Malvezzi, M. Patrini, V. Mulloni, and L. Pavesic, Bulk and surface contributions to second-order susceptibility in crystalline and porous silicon by second-harmonic generation, Surf. Sci. \textbf{481}, 105 (2011).
     %
    \bibitem[Corvi(1986)]{Corvi} M. Corvi and W. Schaich, Hydrodynamic-model calculation of second-harmonic generation at a metal surface, Phys. Rev. B \textbf{33}, 3688 (1986).
    %
    \bibitem[Guyot(1986)]{Guyot} P. Guyot-Sionnest, W. Chen, and Y. R. Shen, General considerations on optical second-harmonic generation from surfaces and interfaces, Phys. Rev. B \textbf{33}, 8254 (1986).
    %
    \bibitem[Somerville(2015)]{Somerville} W. R. C. Somerville, B. Auguie, and E. C. Le Ru, Accurate and convergent \textit{T}-matrix calculations of light scattering by spheroids, J. Quant. Spectrosc. Radiat. Transfer \textbf{160}, 29 (2015).
    %
    \bibitem{COMSOL}COMSOL Multiphysics\textsuperscript{\textregistered}, www.comsol.com.
    \bibitem[Hsu(2016)]{Hsu} C. W. Hsu, B. Zhen, A. D. Stone, J. D. Joannopoulos, and M. Soljacic, Bound states in the continuum, Nat. Rev. Mater. \textbf{1}, 16048 (2016).
    \bibitem[Hsu(2016)]{Wang} J. T. Wang, P. Tonkaev, K. Koshelev, F. Lai, S. Kruk, Q. Song, Y. Kivshar, and N. C. Panoiu, Resonantly enhanced second- and third-harmonic generation in dielectric nonlinear metasurfaces, Opto-Electron. Adv. \textbf{7}, 230186 (2024).

\end{thebibliography}
%\bibliographystyle{aipauth4-1}

\end{document}